\documentclass[12pt]{article}
\usepackage{eufrak}
\usepackage{latexsym}
\usepackage[mathscr]{eucal}

\def\be{\begin{equation}}
\def\ee{\end{equation}}
\def\bea{\begin{eqnarray}}
\def\eea{\end{eqnarray}}
\def\beqa{\begin{eqnarray}}
\def\eeqa{\end{eqnarray}}
\def\beq{\begin{equation}}
\def\eeq{\end{equation}}
\def\beqal{\begin{eqnarray}\label}
\def\beql{\begin{equation}\label}

\def\R{\mbox{\rm I\kern-.18em R}}
\def\Rq{\R^4}
\def\P{\mbox{\rm I\kern-.18em P}}
\def\uno{\mbox{1 \kern-.59em {\rm l}}}
\def\Ds{\ {\big / \kern-.70em D}}
\def\ds{\big / \kern-.90em {\ \p} }

\def\cDs{\ {\big / \kern-.70em {\cal D}}}
\def\Z{{Z \kern-.45em Z}}
\def\Q{{\kern .1em {\raise .47ex \hbox{$\scriptscriptstyle |$}}
\kern -.35em {\rm Q}}}

\def\Tr{\mbox{\rm Tr}}

\def\p{\partial}

\def\al{\alpha}

\def\m{\mu}
\def\n{\nu}

\def\r{\rho}
\def\s{\sigma}

\def\L{\Lambda}

\def\ie{{\it i.e.}}

\def\1{\dot{1}}
\def\2{\dot{2}}
\def\ccc{\raisebox{+.3ex}{$\stackrel{\scriptstyle \leftrightarrow}{\cal S}$}}

\def\ie{{\it i.e. }}

\font\mybb=msbm10 at 12pt
\font\mybbb=msbm10 at 8pt
\def\bbb#1{\hbox{\mybbb#1}}
\def\bb#1{\hbox{\mybb#1}}
\def\Z {\bb{Z}}
\def\C {\bb{C}}

\def\bC {\bbb{C}}
\def\bR {\bbb{R}}
\def\bZ {\bbb {Z}}
\def\real{{\bb{R}}}

\def\Ker{\mbox{Ker}}

\begin{document}
\begin{titlepage}
\begin{flushright}
{ROM2F/2000/04}\\
{DFPD00/TH/09}
\end{flushright}
\begin{center}
 
{\large \sc Instanton Calculus,  Topological Field Theories and
\\
$N=2$ Super Yang--Mills Theories
}\\ 
  
\vspace{0.2cm}
{\sc Diego Bellisai}\\
{\sl Dipartimento di Fisica ``G. Galilei'', Universit\`a di Padova\\
and I.N.F.N. Sezione di Padova,
Via Marzolo 8, 35131 Padova, Italy}
\vskip 0.2cm
{\sc Francesco Fucito,}\\
{\sl I.N.F.N. Sezione di Roma II,   
Via della Ricerca Scientifica, 00133 Roma, Italy}\\
\vskip 0.2cm
{\sc Alessandro Tanzini,}\\
{\sl Dipartimento di Fisica, 
Universit\`a di Roma  ``Tor Vergata"}\\
and
\\
{\sc  Gabriele Travaglini}\\
{\sl I.N.F.N. Sezione di Roma II,   
Via della Ricerca Scientifica, 00133 Roma, Italy}
\vskip 0.5cm
{\large \bf ABSTRACT}\\
\end{center}
{The results obtained by Seiberg and Witten for the low--energy
Wilsonian effective actions  of $N=2$ supersymmetric theories 
with gauge group $SU(2)$  are in agreement with  instanton computations
carried out for winding numbers one and two. 
This suggests that the instanton saddle point saturates 
the non--perturbative contribution to the functional integral. 
A natural framework in which corrections to this  approximation 
are absent is given  by the topological field theory built out of the
$N=2$ Super Yang--Mills theory. 
After extending the standard construction of 
the Topological Yang--Mills theory to encompass the case of a  
non--vanishing vacuum expectation value for the scalar field,  
a BRST transformation  is defined (as a supersymmetry plus a gauge variation), 
which  on  the instanton moduli space is the exterior derivative. 
The topological field theory approach makes 
the so--called ``constrained instanton" configurations 
and the instanton measure   arise in a natural way.  
As a consequence, instanton--dominated Green's functions in 
$N=2$ Super Yang--Mills can be equivalently computed 
either using the constrained instanton method 
or making reference to the topological twisted version of the theory. 
We explicitly compute the instanton  measure  
and the contribution to $u=\left< \Tr \phi^2 \right>$ 
for winding numbers one and two. 
We then show that each non--perturbative 
contribution to the $N=2$ low--energy effective action 
can be written as  the  integral of a total  
derivative of a  function of the instanton moduli.
Only instanton configurations of  zero conformal size contribute to 
this result. 
Finally, the $8k$--dimensional instanton moduli space  is built 
using the hyperk\"ahler quotient procedure, which 
clarifies the geometrical meaning of our approach.}
                            
\vfill
\end{titlepage}
\addtolength{\baselineskip}{0.3\baselineskip} 
\setcounter{section}{0}
\section{Introduction}                   

Our understanding of the non--perturbative sector of field
and string theories has greatly progressed in recent
times. In \cite{sw}, for the first time, the entire non--perturbative
contribution to the holomorphic part of the 
Wilsonian effective action was computed
for $N=2$ globally supersymmetric (SUSY) theories with gauge group $SU(2)$,
using ans\"atze dictated by physical intuitions.
A few years later, 
a better understanding of non--perturbative configurations
in string theory led to the conjecture that 
certain IIB string theory correlators on an  
$AdS_5\times S^5$ background are related to 
Green's functions of composite operators of an 
$N=4$  $SU(N_c )$ Super Yang--Mills
(SYM) theory in four dimensions in the large $N_c$ limit
\cite{mal}. 
Although supported by many arguments,
these remarkable results remain conjectures and a clear mathematical 
proof seems to be out of reach at the moment. In our opinion
this state of affairs is mainly due to the lack of adequate computational 
tools
in the non--perturbative region. To the extent of our knowledge, the 
only way to perform computations in this regime 
in SUSY theories and from first principles 
 is via multi--instanton calculus.
Using this tool,  many partial checks 
have been performed on these conjectures,  
both in   $N=2$ and $N=4$ SUSY gauge 
theories \cite{fp, DKM, ft, bkgr, DKM3}.
The limits on these computations come from the exploding amount of algebraic
manipulations to be performed and from the lack
of an explicit parametrization of instantons of winding number greater
than two \cite{adhm}. 
In order to develop
new computational tools that might allow an extension 
to arbitrary winding  number,  we revisit 
instanton computations for $N=2$ in the light of the topological theory 
built out of  $N=2$ SYM,
{\it i.e.} the so--called Topological Yang--Mills 
theory (TYM) \cite{witten}.%
\footnote{This is somewhat different from previous work
\cite{mw} in which the spectral curves which describe the moduli space of 
vacua for $N=2$ theories with various gauge groups were put in relation with
integrable systems which, in turn, are related to 2--dimensional
topological field theories. 
The study of the relationship   between this approach and the one we present
here goes beyond the scope of this paper.} 

That the TYM might play an important r\^{o}le
in instanton computations became apparent with the results of 
\cite{fp,DKM, ft}. The agreement of these computations with the
results of \cite{sw} pointed out that  instantons 
saturate all the non--perturbative contributions to the
$N=2$ SYM low--energy effective action, and that 
the saddle point expansion
around the classical solution does not receive any perturbative
correction. This situation  seems to be related to a sort of 
localization theorem \cite{at}.
In this respect $N=2$ stands as an isolated case. Its effective action
can be separated into a holomorphic  part 
(which encodes the geometry of the quantum moduli space of 
vacua) and a non--holomorphic one. 
In \cite{sei} a powerful non--renormalization theorem
for the former   was found. 
Subsequently, 
the  quantum holomorphic piece was exactly determined \cite{sw}, and   its 
non--perturbative part 
checked against instanton calculations \cite{fp,DKM, ft}.
It is this kind of contributions  that we 
claim can also be computed from a closely related 
topological field theory. The story for the 
non--holomorphic terms is completely different. Indeed, the 
leading instanton contribution to the higher--derivative terms 
in the effective action  
gets perturbative corrections \cite{bfmt},
in a way similar to 
what occurs in the $N=4$ SUSY theory, 
in which the non--perturbative contributions 
to all the relevant correlators get also 
perturbatively corrected, as a consequence of the  
vanishing of the $R$--symmetry anomaly. 

From now on we will focus on the $N=2$ case. First of all, 
we generalize the standard approach to  TYM \cite{witten}
to encompass the case in which the scalar field  acquires 
a non--vanishing vacuum expectation value (v.e.v.), and
define a BRST operator on field space,  which 
on the moduli space  $\mathscr{M}^+$ ($\mathscr{M}^-$) of 
(anti--)self--dual gauge connections acts as the exterior derivative. 
This allows us to show that the computation of the relevant  Green's functions 
boils down to integrating differential forms on 
$\mathscr{M}^+$ ($\mathscr{M}^-$).
More precisely,  after integrating out the quantum fluctuations 
one is left with a theory 
living on the instanton moduli space.  To describe this space 
we will make use of the ADHM construction \cite{adhm}. 

The TYM framework allows for a better understanding of the geometry 
underlying the computation of correlators of observables, 
and casts also new light on old problems concerning 
instanton calculus. On one hand we will learn that the 
BRST operator built with a zero
v.e.v. for the scalar and that  obtained with a non--zero v.e.v. 
cannot be smoothly deformed one into the other. Thus, it does not
make sense trying to match computations performed in these two
different regimes. 
On the other hand,  in the case of non--zero scalar v.e.v.,
the twisted formulation naturally leads to  the construction
of the so--called ``constrained instanton'' field
configurations \cite{af,nsvz}
thus giving a firmer basis to this approach.
The Ward identities associated to the scalar supersymmetry 
transformations show that the 
constrained instanton computational  method actually gives  
the correct result for the Green's functions
of physical observables.
This in turn implies that, as argued before,  
instantons saturate the non--perturbative 
contribution  to the relevant Green's functions, 
and shows  how the  non--renormalization  theorem of \cite{sei} 
explicitly works in the context of instanton calculus. 

It is worth  remarking that 
in the geometrical approach outlined here, 
the instanton measure  for the $N=2$ SYM theory  
({\it i.e.} the integration measure over the moduli, 
or collective coordinates) emerges in a very natural way, 
without resorting to any intricate zero--mode calculation 
as in previous approaches.
In the same vein, we derive with purely algebraic methods
an explicit realization of the BRST algebra on instanton moduli space.
This derivation is deeply related to the 
construction of this space as a hyperk\"ahler quotient \cite{hklr}, 
which we study in the last section. 

In the case of non--zero scalar v.e.v., the instanton action, 
{\it i.e.} the $N=2$ SYM action functional computed on the zero--modes 
(in our picture these are the field configurations 
onto which the action functional of TYM projects)
can be interpreted as the commutator of the BRST charge with an 
appropriate function. This leads to the possibility  
of writing the correlators of physical observables 
as integrals of total differentials  on $\mathscr{M}^+$.
The circumstance that these Green's functions  can   be  
computed in principle on the boundary of $\mathscr{M}^+$ may greatly help in 
computations,  since instantons on $\partial\mathscr{M}^+$ obey a kind
of dilute gas approximation,  as we will explain 
in subsec.\,\ref{cracchis}. This might lead to 
recursion relations of the type found  in  
\cite{mat,edel,hoker}, and simplify instanton calculations.
We also explore how the geometrical approach described here 
works  in the case of vanishing v.e.v., and apply these ideas to 
the computation of correlators in Witten's topological field theory. 

To avoid misunderstandings, we stress that
the fact that certain correlators of  $N=2$ SYM can be calculated 
using the formalism of the TYM does not mean that 
the former is a topological theory: 
$N=2$ SYM is a ``physical'' theory with
its own degrees of freedom and a running coupling constant. In fact, 
TYM is formally derived from $N=2$ SYM by the twisting procedure
which, in flat space, turns out to be just a variable redefinition. 
However,  promoting the scalar supersymmetry generator  
present in the twisted $N=2$  algebra
to a BRST charge
implies great changes in the physical interpretation of the theory; 
some fields become ghosts and their engineering dimensions
change \cite{anfr}. 
TYM theory deserves its name topological because it is a theory
with zero degrees of freedom, whose correlators 
can be related  to topological invariants 
of the four dimensional manifold on which the theory
lives. Also  in SUSY gauge theories a class of position--independent 
correlators exists \cite{nsvz,akmrv}. 
One realizes that 
a subset of correlators of $N=2$ SYM coincides with 
a subset of the observables defined in TYM over $\Rq$: 
as a consequence, 
these Green's functions  can be computed in either theory, according
to one's preferences.
 
Summarizing, we believe this approach  
provides us with  a natural and simplifying  framework 
for  investigating the 
non--perturbative dynamics of  SUSY gauge theories. 
An abbreviated account of part of the results described here 
was presented  in  \cite{bftt}. 

This paper is organized  as follows. 
In subsec.\,2.1 we recall some 
basics of topological field theories with vanishing  v.e.v. for
the scalar field. We derive  the set of identities 
which define a BRST operator
and show that the functional integration projects the fields onto 
the subspace of the zero--modes of the relevant kinetic operators
in the instanton background. 
In subsec.\,2.2 we generalize
this discussion to the case of a non--vanishing v.e.v. and 
clarify the relationship between our approach and the 
constrained instanton computational method. 
In subsec.\,3.1,  
after an introduction  to the ADHM construction of 
instantons, we write the solutions to the equations of motion derived 
from the TYM action. 
In subsec.\,3.2 we use
the results of the previous subsection and the  identities  associated to the 
BRST symmetry  to find how
the BRST charge acts on the relevant quantities defined in the 
ADHM construction 
({\it e.g.} the instanton moduli)
in the absence of a v.e.v. for the scalar field, and in subsec.\,3.3. 
in the case of a 
non--vanishing v.e.v. We present in subsec.\,3.4 a purely 
algebraic (and independent) derivation of the BRST algebra on 
instanton moduli space 
and  of the solutions to the equations of motion (which were obtained 
 in the  previous subsections).
In sec.\,4 we  discuss how to compute  instanton--dominated Green's
functions using the formalism 
we have developed. It is important to understand how in  our approach 
the instanton measure arises. This crucial issue is discussed in  
subsec.\,4.1, where we also study in detail the cases of winding number 
one and two. 
In  subsec.\,4.2 we compute the multi--instanton action (which is 
non--zero when 
the scalar field acquires a non--vanishing v.e.v.)
and show that  it can be written as a BRST--exact  quantity. 
Sec.\,5 is devoted to the calculation of $u=\left< \Tr \phi^2 \right>$, 
the gauge invariant quantity which parametrizes 
the moduli space of quantum vacua of the $N=2$ SYM theory  (from  
which one can obtain  the Seiberg--Witten low--energy Wilsonian action
using Matone's relation \cite{mat}).
First, we find a general expression in our framework for 
the $k$--instanton contribution 
to $u$. Then, 
on one hand,  in subsec.\,5.1 and 5.2,  we compute 
$\left< \Tr \phi^2 \right>$ 
in the bulk of $\mathscr{M}^+$ 
for winding numbers $k=1, 2$; on the other hand, in subsec.\,5.3,  
using the observation of subsec.\,4.2,  we show  that the contribution 
to $u$ can be written as 
a total derivative integrated on the moduli space of  instantons. 
This suggests the interesting possibility of computing it 
directly on the boundary of $\mathscr{M}^+$; we 
explicitly check this  in a $k=1$ computation,  getting the correct result.
In sec.\,6 we consider the case of a vanishing v.e.v. 
(to which our formalism also applies),  and compute
$\left< \Tr \phi^2\Tr \phi^2 \right>$ 
for winding number $k=1$ both in the bulk and on  
the boundary of the instanton moduli space. 
Finally, in sec.\,7 we construct the metric on the $8k$--dimensional  
moduli space of self--dual gauge connections for  winding number $k=2$ 
following the aforementioned hyperk\"ahler quotient procedure.
\section{Topological Yang--Mills Theory}
\setcounter{equation}{0}
\label{startrek}

It is well known that, if the generators
of the rotation group of $\Rq$ are redefined in a suitably twisted fashion, 
the $N=2$ SYM
theory gives rise to the TYM theory considered in \cite{witten}. 
A  key feature of 
the twisted theory is the presence of a scalar fermionic symmetry $Q$, 
which is still an invariance  of the theory  when this  is formulated 
on a generic (differentiable) four--manifold $M$.
This scalar symmetry will play a major r\^{o}le, for the following reasons.
First, the correlation functions
of the physical observables of the theory 
(the cohomology classes of $Q$)
are independent of the metric on $M$ 
by virtue of the  Ward identities associated to $Q$ 
\cite{witten}.
This also implies that these functions must be independent of
the positions of the operatorial insertions.%
\footnote{That in some SUSY gauge theories
there exists a class of position--independent correlators 
was observed in \cite{akmrv}.} 
Moreover, the same Ward identities 
entail that certain Green's functions can be computed exactly in the 
semiclassical limit; this is why instantons come into play. 
Finally, when one modifies  
the scalar  supersymmetry charge $Q$ 
to make it nilpotent, the resulting (BRST) operator 
acts as the exterior derivative on the anti--instanton moduli 
space $\mathscr{M}^-$ \cite{witten2}. 
As we will see, functional integration reduces to 
integrating differential forms on $\mathscr{M}^-$
(this is what we call the localization procedure).
We will later show that the Green's functions of the observables 
can be written as  integrals of total derivatives on 
$\mathscr{M}^-$. 

Let us first recall  how the twisting  operation works  \cite{witten}. 
The global symmetry group of the $N=2$ SUSY theory in flat space is
\beq 
SU(2)_L\times SU(2)_R\times SU(2)_A\times U(1)_R \ \ ,
\label{gr} 
\eeq 
where the first two factors represent the
Euclidean Lorentz group ({\it i.e.}
the rotation group of $\Rq$), while $SU(2)_A$ is  the
automorphism group of the $N=2$ supersymmetry algebra and  $U(1)_R$ is the
usual $R$--symmetry. The twist  consists in
replacing  one of the factors of the rotation group, say for
definiteness $SU(2)_R$, with a diagonal subgroup  $SU(2)_R^\prime$
of $SU(2)_R \times SU(2)_A$.
The symmetry group of the twisted theory is then
\beq 
SU(2)_L\times
SU(2)_R^\prime\times U(1)_R \ \ . 
\label{tw-gr}
\eeq 
With respect to
the twisted group, the SUSY charges decompose as a scalar $Q$,
a self--dual antisymmetric tensor $Q_{\mu\nu}$ and a vector $Q_\mu$: 
\beqa
&&\bar{Q}_{\dot\alpha}^{\dot A}\rightarrow Q \oplus Q_{\mu\nu} \ \
, \nonumber\\ 
&&Q_{\alpha}^{\dot A}\rightarrow Q_\mu \ \ .
\label{tw-ch} 
\eeqa 
In particular, the charge $Q$ belongs to the
$(0,0)^1$ representation of (\ref{tw-gr}), while the charges
$Q_\m, Q_{\m\n}$ belong respectively to the $({1\over 2},{1\over
2})^{-1}$ and $(0,1)^1$ representations.%
\footnote{With the upper index we denote the $R$--symmetry charge.}
In the twisted theory it is natural to 
redefine the fields of $N=2$ SYM  as 
\beqa 
&&A_\mu \rightarrow A_\mu
\ \ , \nonumber\\ 
&&\bar{\lambda}_{\dot\alpha}^{\dot A}\rightarrow
\eta \oplus \chi_{\mu\nu} \ \ , \nonumber\\
&&\lambda_{\alpha}^{\dot A} \rightarrow \psi_\mu \ \ , \nonumber\\
&&\phi \rightarrow \phi \ \ .
\label{tw-fi} 
\eeqa 
The anticommuting fields $\eta,\chi_{\m\n},\psi_{\m}$ are respectively
a scalar, a self--dual antisymmetric tensor  and a vector, belonging to the
$(0,0)^{-1}$, $(0,1)^{-1}$ and $({1\over 2}, {1\over 2})^1$
representations  of the twisted group; 
the gauge field $A_\m$ and the scalar field $\phi$ belong
respectively to the $({1\over 2},{1\over 2})^0$ and $(0,0)^2$ representation.

In the following we will be mainly interested in the multiplet 
of fields $(A_\mu , \psi_\mu , \phi )$, whose 
transformations under the action of $Q$ read
\beqa 
&&Q A_\mu = \psi_\mu \ \ , \nonumber\\ 
&& Q \psi_\mu = - D_\mu \phi \ \ , \nonumber \\ 
&& Q \phi = 0 \ \ . 
\label{delta} 
\eeqa
These equations imply  that $Q$  is   nilpotent modulo gauge
transformations with parameter $\phi$, since 
\beqa 
&& Q^2 A_\m = - D_\m \phi \ \ , \nonumber\\ 
&& Q^2\psi_\m = - [\phi, \psi_\m] \ \ , \nonumber\\
&& Q^2\phi = 0 \ \ 
\label{delta2}
\eeqa 
(this is analogous to what one comes across in studying 
the supersymmetry algebra  in the Wess--Zumino gauge).
(\ref{delta2}) allows us to regard $Q$ as a
BRST--like charge.  
To this end, let us  
assign  $Q$ a ghost number equal to 1; this can be done
by simply identifying its $R$--charge with the ghost number.
Accordingly,  the fields of the twisted $N=2$ vector multiplet 
(\ref{tw-fi}) acquire a ghost number equal 
to their respective $R$--charge.
Since the canonical dimension of the BRST operator
is usually taken to be zero, 
we redefine the canonical dimension
of $Q$ to zero. 
The resulting canonical dimensions and ghost numbers of the
fields are summarized in the table below. 

\begin{center}
\begin{tabular}{|c||c|c|c|c|c|c|c|c|c|}
\hline Fields & $A$ & $\psi$ & $\chi$ & $\eta$ & $\phi$ & $\bar\phi$ \\ 
\hline\hline dimension &  1 & 1 & 2 & 2 & 0 & 2 \\ 
\hline ghost \# & 0 & 1 & -1 & -1 & 2 & -2 \\ 
\hline
\end{tabular}\\
\end{center}
\vskip 0.5cm

At this point we need to distinguish
the case in which the scalar v.e.v. vanishes 
from that in which it is non--vanishing. 
In the next subsection we will focus on the former situation; 
the latter requires a more detailed discussion, and will be 
studied separately in subsec.\,\ref{venanzio}.
\subsection{Case I: Zero  Vacuum Expectation Value for the Scalar Field}
\label{vivenzio} 
As (\ref{delta2}) shows,  $Q$ is not nilpotent. A strictly nilpotent
BRST charge can be obtained from $Q$ by 
introducing a  generalized BRST operator $s$ including 
both the gauge symmetry and
the scalar supersymmetry of the theory  \cite{bs},
\beq 
s = s_g + Q  \ \ . 
\label{brst-op} 
\eeq 
$s_g$ is the usual BRST
operator associated to the gauge symmetry,
\beqa 
&&s_g A = - Dc \ \ , \nonumber\\ 
&& s_g \psi = - [c,\psi] \ \ , \nonumber\\ 
&& s_g \phi = - [c,\phi] \ \ , \nonumber\\ 
&& s_g c = - {1\over 2} [c,c] \ \ ,
\label{gauge} 
\eeqa
the ghost number and canonical dimension of the ghost field $c$ being
respectively one and zero.  
The 1--form  $A = A_\m dx^\m$ is the gauge connection, 
with curvature $F = 1/2 F_{\m\n} dx^\m dx^\n = dA + AA$; 
$\psi = \psi_\m dx^\m$ is an anticommuting 1--form, 
and $D$ is the covariant exterior derivative on the manifold $M$. 
The action of $Q$ on the ghost field $c$ is obtained by requiring that 
$s^2=0$, and turns out 
to be simply 
\be
\label{acci}
Qc=\phi
\ \ .
\ee
(\ref{delta}), (\ref{gauge})  and (\ref{acci})   thus lead
to the following BRST identities \cite{bs}:
\beqa
\label{BRST} 
&&sA=\psi-Dc\ \
,\nonumber\\ &&s\psi=-[c,\psi]-D\phi\ \ ,\nonumber\\
&&s\phi=-[c,\phi]\ \ ,\nonumber\\ 
&&sc=-{1\over 2}[c,c]+\phi\ \ ,
\eeqa
The algebra (\ref{BRST}) can be read as the definition and the
Bianchi identities for the curvature 
\beq
\widehat{F}=F+\psi+\phi
\label{effepiupsi}
\eeq
of the connection 
\beq
\widehat{A}=A+c 
\label{apiuci}
\eeq
of the universal bundle $P\times {\cal A}/{\cal G}$, 
where $P, {\cal A}, {\cal G}$ are respectively the
principal bundle over $M$, the space of connections and the group
of gauge transformations.
The exterior derivative on the manifold 
$M\times {\cal A}/{\cal G}$ is given by \cite{bs}
\beq
\widehat{d}=d+s \ \ .
\label{dipiuesse}
\eeq
Notice that from the last of (\ref{BRST}) we learn that the 
scalar  field $\phi$ can be seen as the curvature of the
connection $c$.

We now come to define the observables of the TYM theory;
these  are given by the elements of the equivariant
cohomology of $s$ \cite{stora}, which satisfy 
the descent equations
\beqa
&& s\ {1\over 2} \Tr F^2 = - d\ \Tr F \psi  \ \ , \nonumber\\
&& s\ \Tr F \psi =  
- d\ \Tr \Bigl(\phi F + {1\over 2} \psi^2\Bigr) \ \ ,\nonumber\\
&& 
s\ \Tr\Bigl(\phi F + {1\over 2} \psi^2\Bigr) = 
-d\ \Tr \phi \psi \ \ ,\nonumber\\
&& s\ \Tr \phi \psi = - {1\over 2} d\ \Tr \phi^2 \ \ , \nonumber\\ 
&& s\ {1\over 2}\Tr \phi^2 = 0 \ \ .
\label{desc-eq}
\eeqa
(\ref{desc-eq}) allows one to 
build local functions of the fields which are BRST 
invariant modulo $d$--exact terms;  
the simplest example of a physical observable is 
the gauge invariant polynomial $\Tr \phi^2$, 
as the last of  (\ref{desc-eq}) shows.
We will see in sec.\,\ref{brstinst} that  this geometrical approach
provides us with  an operative tool which  allows us to compute 
the Green's functions  of observables 
starting only from the knowledge of 
the universal connection 
(\ref{apiuci}),  in particular without solving any equation of motion.

As shown in \cite{bs}, a TYM action can be interpreted as a pure 
gauge--fixing term,  
\beq 
S_{\rm TYM} = 2 \int d^4x~s \Tr \Psi \ \ ,
\label{s-tym}
\eeq
where the gauge--fixing fermion is chosen to be
\beq
\Psi = \chi^{\m\n} F^+_{\m\n} - D^\m\bar\phi\psi_\mu + 
\bar c \partial^\m A_\m \ \ , 
\label{gf-ferm}
\eeq 
and 
\beq 
F_{\m\n}^+ = {1\over 2}
\Bigl(F_{\m\n} + {1\over 2}
\epsilon_{\m\n\r\s} F^{\r\s}\Bigr) 
\nonumber 
\eeq 
is the
self--dual component of the field strength $F_{\m\n}$. 
The anti--fields $\chi_{\m\n}$, $\bar\phi$ and $\bar c$ transform 
under the BRST symmetry  as 
\beqa 
&& s\chi_{\m\n} = B_{\m\n}\ \ , \nonumber\\
&& s \bar\phi = \eta\ \ , \nonumber\\
&& s \bar c = b\ \ , 
\label{anti-fi}
\eeqa 
while the
Lagrange multipliers $B_{\m\n}$, $\eta$ and $b$ as 
\beqa 
&& s B_{\m\n} = 0 \ \ ,\nonumber\\ 
&& s\eta = 0 \ \ , \nonumber \\ 
&& s b = 0 \ \ . 
\label{l-mult}
\eeqa
The anti--field $\bar c$ has ghost number $-1$ and dimension $2$,
while the Lagrange multipliers ($B_{\m\n}$, $b$) have ghost number $0$ and
dimension $2$.
Acting with the BRST operator $s$ in (\ref{s-tym}), we obtain
the following explicit form for the TYM action 
\beqa
S_{\rm TYM} &=& 2\int d^4x \ \Tr \Bigl[
 B^{\m\n}F_{\m\n}^+ - \chi^{\m\n}(D_{[\m}\psi_{\n]})^+ 
+ \eta  D^\m \psi_\m + \nonumber\\
&& - \bar\phi ( D^2 \phi - [\psi^\m,\psi_\m] ) 
+ b \partial ^\m A_\m + \nonumber\\
&& + \chi^{\m\n} [c,F_{\m\n}^+] - \bar\phi [c,D^\mu\psi_\mu] - 
\bar c s(\partial^\m A_\m) \Bigr] \ \ ,
\label{tym-act}
\eeqa
where 
\beq
(D_{[\m}\psi_{\n]})^+ = {1\over 4} \Bigl( D_\m \psi_\n - D_\n\psi_\m + 
\epsilon_{\m\n\r\s} D^\r\psi^\s \Bigr)
\eeq
is the self--dual component of the tensor $D_{[\m}\psi_{\n]}$.
(\ref{tym-act}) is obtained integrating  
by parts the term in $\bar\phi$ of (\ref{gf-ferm});   
the corresponding surface term vanishes, 
since in this 
subsection we limit ourselves to study 
the case in which all 
the fields have trivial boundary 
conditions.

The main property of the action (\ref{tym-act}) is that 
{\it it  localizes the fields in the algebra (\ref{BRST}) onto certain 
sections of the universal bundle.} 
In particular,
functional integration over the fields 
$B^{\m\n}$ and ($\chi^{\m\n}$, $\eta$) 
in the first line of (\ref{tym-act}) leads 
respectively to 
\beq
F_{\m\n}^+ = 0 \ \ , 
\label{mod}
\eeq
which implies that $A$ is an anti--self--dual gauge connection,
and
\beqa
&& (D_{[\m}\psi_{\n]})^+ = 0 \ \ , \nonumber\\
&& D^\m\psi_\m = 0 \ \ , 
\label{tmod}
\eeqa
which entails that $\psi$ is an element of the tangent bundle 
$T_A\mathscr{M}^-$.
In turn,  
functional integration over the anti--field $\bar\phi$ leads to 
the equation
\beq
D^2\phi = [\psi^\m, \psi_\m] 
\label{phi-eq}
\eeq
for the field $\phi$, while integration on the Lagrange multiplier
$b$ imposes the usual transversality condition 
\beq
\partial^\m A_\m = 0 \ \ .
\label{transv}
\eeq
By plugging the expression for $\psi$ deduced  from the first equation in 
(\ref{BRST}) into the transversality condition $D^\m\psi_\m=0$, we get
\beq
D^2 c = - D^\m (s A_\m) \ \ ,
\label{c-eq}
\eeq
which determines the ghost field $c$.
Two observations are in order. 
First, let us remark that the first equation in 
(\ref{BRST}) is an old acquaintance \cite{th, bern}. It is in fact very 
well known that differentiating the gauge connection 
with respect to collective coordinates ($s A$)  fails to give 
a transverse zero--mode ($\psi$).
To ensure the correct gauge condition,  the addition of a gauge
transformation  ($Dc$) is needed: 
(\ref{c-eq}) just determines  this gauge transformation. 
A suitable framework for multi--instanton calculations 
is given by the ADHM construction.
It is interesting to discover that  this construction, 
together with the TYM  formalism, 
allows one to write  the universal connection (\ref{apiuci})  
(and consequently  the ghost $c$ and the gauge transformation $Dc$) 
in a very natural and straightforward  way \cite{anselmi}. 
We will focus on  this aspect in sec.\,\ref{brstinst}.
Finally, 
the terms in the last line of (\ref{tym-act}) vanish 
due to the conditions
(\ref{mod}), (\ref{tmod}) and (\ref{transv}).

Summarizing, we have seen that after functional integration on 
the anti--fields and the Lagrange 
multipliers,
we are left with an integration on the space of anti--self--dual gauge 
connections 
$\mathscr{M}^-$ and
its tangent bundle $T_A\mathscr{M}^-$, with a functional measure equal to
1, since the action 
$S_{\rm TYM}$ vanishes on the field  subspace identified by 
the (zero--mode) equations (\ref{mod})--(\ref{c-eq}). 

Notice that in this approach the functional integral is performed 
exactly, since the gauge--fixing fermion 
(\ref{gf-ferm}) is linear in the antifields,   
and there are no perturbative corrections. 
It is important to remark that  
the action obtained by twisting the $N=2$ SYM theory  
({\it i.e.} the action of Witten's topological field theory \cite{witten})  
actually differs from (\ref{tym-act}) 
 by some extra terms, which spoil the
linearity of $S_{\rm TYM}$ in the anti--fields. However, 
as we will show below,  they are BRST--exact terms   
corresponding to a continuous
deformation of the gauge--fixing
\beq
\label{quiquoqua}
S_{N=2}= S_{\rm TYM} + s {\cal V}
\ \ ;
\eeq
the v.e.v. of an $s$--closed operator ${\cal O}$,
{\it i.e.} such that $s{\cal O}=0$, 
is controlled by  
the Ward identity \cite{witten}
(in the following $[\delta \varphi]$ is  shorthand for 
the integration measure)
\beqa
< {\cal O }>_{S + s {\cal V}}  &\equiv & 
\int [\delta \varphi]\ e^{-(S + s{\cal V })}
{\cal O } =  \int [\delta \varphi]\ e^{-S }
{\cal O } (1 - s {\cal V} + \cdots ) 
\nonumber \\
&=& 
< {\cal O }>_{S} - < {\cal O } s{\cal V}>_S  + \cdots = 
< {\cal O }>_{S} - < s( {\cal O } {\cal V}) >_S + \cdots = 
\nonumber \\
&=&
< {\cal O }>_{S}
\ \ ,
\label{quiquo}
\eeqa 
the last equality following  from the fact that
 the v.e.v. of  an $s$--exact operator 
${\cal P} =  s {\cal Q}$ vanishes  if ${\cal Q}$
is globally defined.%
\footnote{In this respect we remark that $\Tr\phi^2$ is not globally
defined. This fact plays an important r\^ole in breaking $N=2$ SUSY into 
$N=1$ \cite{witten3}.}
This means that the action (\ref{tym-act}) and 
the twisted $N=2$ SYM action are completely  equivalent, in the sense that  
{\it 
the Green's functions of  $s$--closed operators 
can be computed using  any  one of them obtaining the same result}. 
We now show  that 
the Lagrangian obtained by twisting the
$N=2$ SYM theory can be derived by modifying the gauge--fixing
fermion  (\ref{gf-ferm}), thus proving that 
the twisted version of $N=2$ SYM action introduced in \cite{witten}
and the TYM action  (\ref{tym-act}) differ only by BRST--exact 
terms \cite{bs}. 
To this end, let us consider the modified gauge--fixing fermion  
\beq
\Psi^{(\al)} = 
\chi^{\m\n}\bigl(F^+_{\m\n} -
{\al\over 2} B_{\m\n}\bigr) - D^\m\bar\phi\psi_\mu + 
\bar c \partial^\m A_\m \ \ , 
\label{al-gf-ferm}
\eeq 
where $\al$ is a gauge--fixing parameter;   upon
functional integration over the
Lagrange multiplier $B_{\m\n}$,  one gets 
\beqa
S_{\rm TYM}^{(\al)} &=& 2 \int d^4x \ \Tr \Bigl[
 {1\over {2\al}}F^{+\m\n}F_{\m\n}^+ - \chi^{\m\n}(D_{[\m}\psi_{\n]})^+ 
+ \eta D^\m \psi_\m + \nonumber\\
&& - \bar\phi ( D^2 \phi - [\psi^\m,\psi_\m] ) 
+ b \partial ^\m A_\m + \nonumber\\
&& + \chi^{\m\n} [c,F_{\m\n}^+] - \bar\phi [c,D^\mu\psi_\mu] - 
\bar c s(\partial^\m A_\m) \Bigr] \ \ .
\label{al-tym-act}
\eeqa
This functional leads to the same equations of motion of the $N=2$ SYM theory.

Three observations are in order. First, 
notice that the partition function  (or more generally the Green's functions) 
defined by (\ref{al-tym-act}) 
has to be studied using the usual saddle point techniques due to the 
presence of the kinetic term for the gauge field.
As discussed in \cite{witten}, supersymmetry
ensures the cancellation of bosonic and fermionic determinants arising 
when integrating out the non--zero modes \cite{divecchia} and the resulting 
integration is over $\mathscr{M}^-$, which is the same result obtained
from the use of the action (\ref{tym-act}).
Second, the renormalization group invariant scale  
(which multiplies the Green's functions computed in 
the conventional supersymmetric theory) does not appear. 
In fact the divergent term of the one--loop effective action
is proportional to $(F^+)^2$ which is obviously vanishing
on $\mathscr{M}^-$ \cite{birm}.

Last, and most importantly, let us observe that 
the equivalence between $S_{\rm TYM}^{(\al)}$, Eq. (\ref{al-tym-act}), and
$S_{\rm TYM}$, Eq. (\ref{tym-act}), 
in the computation 
of correlators of observables  is not  surprising, since they 
cannot depend on the
choice of the gauge parameter $\al$;  therefore nothing prevents us 
from choosing directly $\al=0$.
As we will discuss in sec.\,{\ref{venanzio}, in the presence of a
non--trivial v.e.v. for
the scalar field, the equivalence of (\ref{tym-act}) to the $N=2$ SYM action
(in the sense previously specified)
makes the configurations of the constrained instanton method emerge 
from  functional integration, without  any approximation procedure.

Let us now sketch the geometrical interpretation 
of the instanton calculus suggested by  
the topological formulation. To begin with, 
the first equation in (\ref{BRST}) together with 
(\ref{tmod}) imply that, as announced, 
 the BRST operator $s$ has an 
intriguing explicit realization on the 
moduli space as the exterior derivative \cite{witten2}. 
Furthermore, once the universal gauge connection
(\ref{apiuci}) is given, the other field configurations $\psi, \phi$ 
are in turn immediately determined 
as components of the universal curvature (\ref{effepiupsi}),
as it will be worked out in detail in sec.\,\ref{brstinst}.
Topological correlators are then built up as differential forms on 
the moduli space,
where the form degree of the fields equals  their ghost number. 
For example, for winding number $k=1$ the top form on the 
(8--dimensional) instanton moduli space is given by
$\Tr \phi^2 (x_1)  \Tr \phi^2  (x_2)$.
We will explicitly compute the corresponding 
Green's function  in sec.\,\ref{vittoria}, 
both with a bulk calculation and with a calculation on the boundary 
of the instanton moduli space, finding the same result.


\subsection{Case II: Non--Zero Vacuum Expectation Value for the Scalar Field}
\label{venanzio}

In this subsection we will extend the construction of the TYM  to encompass
the presence of a non--vanishing v.e.v. for the scalar field.
To this end,  observe first  that a non--zero v.e.v. for $\phi$, 
\beq
\lim_{|x| \rightarrow \infty}\phi = 
v \frac{\sigma_{3}}{2i} \ \ ,
\label{phi-bc}
\eeq
implies the existence of a (non--zero) central charge $Z$ in the SUSY 
algebra.
Then the operator defined in (\ref{brst-op})  is no longer
nilpotent;
instead, it closes on a $U(1)$ central charge transformation 
\beqa
&& (s_g + Q)^2 A = Z A\equiv - D\phi_Z \ \ , \nonumber\\
&& (s_g + Q)^2 \psi = Z \psi\equiv - [\phi_Z, \psi] \ \ , \nonumber\\
&& (s_g + Q)^2 \phi_Z = Z \phi_Z \equiv 0 \ \ ,
\label{Z}
\eeqa
where the scalar field $\phi_Z$ plays the r\^{o}le of
a gauge parameter and satisfies the equation
\beqa
&& D^2 \phi_Z = 0 \ \ , \nonumber\\
&& \lim_{|x| \rightarrow \infty}
\phi_Z  = v \frac{\sigma_{3}}{2i}\ \ .
\label{phi-Z}
\eeqa
Notice that from (\ref{Z}) 
it follows that the central charge $Z$ has ghost number
two and canonical dimension zero. 
We also remind that 
the scalar charge $Q$ commutes with $s_g$, 
while $Z$ commutes with all the other charges by definition. 

From (\ref{Z}) it follows that, in order to ensure the nilpotency property,
the operator (\ref{brst-op}) has to be 
properly extended by including the central charge.
We then define an extended BRST operator \cite{hen} as 
\beq
s = s_g + Q - \lambda Z + {\partial\over{\partial\lambda}} \ \ ,
\label{brstZ-op}
\eeq
where $\lambda$ is a fermionic parameter with ghost number $-1$ and canonical 
dimension zero, such that $\lambda Z$ has the usual quantum numbers of 
a BRST  operator.
It is easy to see that the last term of (\ref{brstZ-op}) is needed to ensure
the nilpotency of $s$; in fact, by using (\ref{Z}) and the property 
$\lambda^2 = 0$,  we get  
\beq
s^2 = ( s_g + Q )^2 - {\partial\over{\partial\lambda}}(\lambda Z) = 0 \ \ .
\label{nilp}
\eeq
If we define the ghost field 
\beq
\L \equiv \lambda \phi_Z\ \ ,
\label{Lambda}
\eeq
with the transformation 
\beq
s \L = \phi_Z - [c,\L] \ \ ,
\label{s-Lambda}
\eeq
we can finally write the resulting BRST algebra as
\beqa
\label{brstZ}
&&sA=\psi-D(c+\L)\ \ ,\nonumber\\
&&s\psi=-[c+\L,\psi]-D\phi\ \ ,\nonumber\\
&&s\phi=-[c+\L,\phi]\ \ ,\nonumber\\
&&s(c+\L)=-{1\over 2}[c+\L,c+\L]+\phi\ \ .
\eeqa
Notice that now the curvature $\phi$ of the ghost $c + \L$ is decomposed
as the sum
\beq
\phi = \phi_Z + \phi_g \ \ ,
\label{phi-dec}
\eeq
where $\phi_g$ is related to the usual gauge transformations
\beqa
&&s c = \phi_g - {1\over 2} [c,c] \ \ , \nonumber\\ 
&& \lim_{|x| \rightarrow \infty} \phi_g = 0 \ \ .
\label{phi-g}
\eeqa
(\ref{Lambda}) entails that the field $\L$, which has ghost
number one and dimension zero, can be seen  as the ghost related to the
central charge symmetry. This fact can be analyzed in more detail by 
considering the equations
(\ref{tmod}) for the $\psi$ field: in the presence of a non--zero central
charge, they no longer have a unique solution, since
any field configuration of the form $\psi - D\tilde{\L}$ satisfies  the
same equations on the anti--instanton background $F_{\m\n}^+ = 0$
\beqa
&& (D_{[\m}\psi_{\n]})^+ - [F_{\m\n}^+,\tilde{\L}] = 
   (D_{[\m}\psi_{\n]})^+ = 0 \ \ , \nonumber\\
&& D^\m\psi_\m - D^2 \tilde{\L} = D^\m\psi_\m = 0 \ \ ,
\label{tmod-Z}
\eeqa
provided that
\beqa
D^2 \tilde{\L} = 0 \ \ ,&& \nonumber\\
\lim_{|x| \rightarrow \infty} \tilde{\L} \neq 0\ \ ;&&
\label{L-eq}
\eeqa
according to (\ref{phi-Z}), 
this identifies  $\tilde{\L}$ as the parameter of a central charge
transformation.
This degeneracy will be removed once the boundary conditions for the
ghost $\tilde{\L}$ are fixed. 
Notice that if the central charge were zero, 
the equation $D^2 \tilde{\L} = 0 $  would have  trivial boundary conditions; 
its  solution would be $\tilde{\L} = 0$,   and the
degeneracy would disappear, as expected.
In our case instead, from (\ref{phi-Z}) and (\ref{Lambda}) 
it follows that $\L$ is a solution
of (\ref{L-eq}) satisfying the boundary condition
\beq
\lim_{|x| \rightarrow \infty} \L = \lambda v \frac{\sigma_{3}}{2i}
\label{L-bc}
\eeq
induced  by the asymptotic behavior of the scalar field $\phi$. 
The new BRST algebra could then be derived  from (\ref{BRST})  by 
just   shifting the ghost $c$  to $c+\L$; 
$c$ is related  to  the usual gauge transformations, whereas  
$\L$ takes into account the new $U(1)$ transformations generated 
by the central charge.

Let us now evaluate the action (\ref{s-tym}).
This can be done by noticing that, 
according to the algebra (\ref{brstZ}), its explicit form 
can be obtained simply by substituting  the ghost $c$ 
in (\ref{tym-act})
with its shifted version $c+\L$.
Unlike the case studied in the previous subsection, 
the action gets now a   contribution
from integrating by parts the term $s \Tr (D^\m\bar\phi\psi_\m)$
which does not vanish due to the non--trivial boundary conditions
(\ref{phi-bc}). 
Explicitly 
\beq
\label{gcr}
s\int\ d^4 x 
\ 2\,\Tr [(D^{\mu}\bar{\phi})\psi_{\mu}]
=s\int\ d^4 x \ 2\, \partial^{\mu}\Tr(\bar{\phi}\psi_{\mu})-
s\int\ d^4 x \ 2\, \Tr (\bar{\phi}D^{\mu}\psi_{\mu})
\ \ .
\eeq
We then get
\beqa
S_{\rm TYM} &=& 
2\int d^4x \ \Tr \Bigl[
 B^{\m\n}F_{\m\n}^+ - \chi^{\m\n}(D_{[\m}\psi_{\n]})^+ 
+ \eta D^\m \psi_\m + \nonumber\\
&& - \bar\phi ( D^2 \phi - [\psi^\m,\psi_\m] ) 
+ b \partial ^\m A_\m + \nonumber\\
&& + \chi^{\m\n} [c+\L,F_{\m\n}^+] - \bar\phi [c+\L,D^\mu\psi_\mu] - 
\bar c s(\partial^\m A_\m) \Bigr] + \nonumber\\
&& -
2 s\int\ d^4 x \ 2\, \Tr (\bar{\phi}D^{\mu}\psi_{\mu})
+ 2 \int d^4x \ \partial^{\mu} s \Tr ( \bar\phi \psi_{\mu} )
\ \ .
\label{ztym-act}
\eeqa
The functional integration 
over the anti--fields and the Lagrange multipliers 
goes as in subsec.\,\ref{vivenzio}, 
and leads  to the same set of 
(zero--mode) equations
\beqa
&& F_{\m\n}^+ = 0 \ \ , \label{a-eq}\\
&& (D_{[\m}\psi_{\n]})^+ = 0 \ \ , \\
&& D^\m\psi_\m = 0 \ \ , \\
&& D^2\phi = [\psi^\m,\psi_\m] \ \ ;
\label{phieq}
\eeqa
the key difference with respect to sec.\,\ref{venanzio} is that now
the scalar field $\phi$ has non--trivial
boundary conditions as per (\ref{phi-bc}). 
It is worth remarking that the 
(zero--mode subspace) configurations  dictated by 
(\ref{a-eq})--(\ref{phieq}),  
together with the boundary condition (\ref{phi-bc}), are 
exactly those which are exploited 
in the context of the constrained instanton method \cite{af,nsvz}
as {\it approximate} solutions to the saddle point equations. 
Instead, as we have explained, in our 
approach such field configurations 
naturally come into play after functional integrating
the anti--fields and the Lagrange multipliers; 
no approximation is involved. 
The Ward identity   (\ref{quiquo}) of  the previous subsection
applies also here, implying the equivalence of the action 
(\ref{ztym-act}) to that of  the $N=2$ SYM theory in computing the 
Green's functions of the  physical observables.
This explains why the constrained instanton method 
gives the correct  result for the calculation of these  correlators.

Once the connection $\widehat{A} = A + c + \L$ is known, 
the BRST identities (\ref{brstZ}) provide us with the 
field configurations of $A$, $\psi$, $\phi$ which solve 
the equations of motion (\ref{a-eq})--(\ref{phieq}).
This possibility,  and the circumstance that the BRST operator
acts on instanton moduli space as the exterior derivative
conspire to  make  it possible  
to explicitly work out  (in the ADHM formalism)
the aforementioned field configurations 
without solving their equations of motion.
In the next section we will  
first guess the ADHM expression for $\widehat{A}$, 
and then show how the procedure outlined here works.

As in the zero v.e.v. case, functional integration is 
performed exactly, and  we are left with an integration 
over $\mathscr{M}^-$ and $T_A\mathscr{M}^-$.  
No perturbative renormalization of the 
physical correlators calculated with TYM is allowed, 
in agreement with the non--renormalization theorem 
of \cite{sei}.
Green's functions are built up as 
differential forms on the moduli space also in this case.
However,  the functional measure is now crucially different
from 1, since the action computed on the zero--mode subspace
gets a non--vanishing contribution
$S_{\rm inst} $ from
the last term of (\ref{ztym-act}), which reads 
\beq
S_{\rm inst} = 2\int d^4x \ \partial^{\mu} s \Tr ( \bar\phi \psi_{\mu} )
\ \ .
\label{DKM}
\eeq  
This in turn implies that $\exp ( - S_{\rm inst} )$ acts as
a generating functional for differential forms on the moduli space. 
This gives rise to
non--trivial correlation functions  which take contribution 
from topological sectors of {\it any} winding number $k$.
The most interesting example  is 
the v.e.v. of the gauge invariant, $s$--exact  operator  
$\Tr \phi^2$, {\it i.e.}
$u(v) = \langle \Tr \phi^2 \rangle$, which  
plays a prominent r\^{o}le in the context of the Seiberg--Witten model. 
In sec.\,\ref{marcella}  we will focus on  this particular Green's function.
First, in (\ref{tamara3}) we will give 
the general expression for the contribution to $u(v)$ 
coming from 
the topological sector of winding number $k$; 
furthermore   in subsec.\,\ref{bob!} and 
\ref{7777} we will perform the computation 
for instanton number equal to one and two respectively. 
Last, in subsec.\,\ref{cracchis} we will illustrate  the possibility of 
computing  $u(v)$ with a calculation
on the boundary of instanton moduli space. To  support this idea we will  
work out the $k=1$ computation explicitly. 

\section{The BRST Algebra on Instanton Moduli Space}
\setcounter{equation}{0}
\label{brstinst}
When restricted to configurations which obey the equations of motion
dictated by the TYM 
action, the BRST algebra gets realized  on  instanton moduli  space. 
In the following we will construct this realization explicitly. 
To this end we will start  by briefly recalling some basic elements 
of the ADHM construction of instantons, which provides us with a 
parametrization of this moduli space. 
This description is given in terms of a redundant set of 
parameters;  we will then focus on its reparametrization symmetries,
which will play a major r\^{o}le in the following. 
Our first goal  will be the construction 
of the BRST algebra on instanton moduli space starting from 
the knowledge of the solutions to (\ref{mod}), (\ref{tmod}), (\ref{phi-eq}), 
(\ref{c-eq}), for a generic winding number, 
which were  found in \cite{adhm,cgt,DKM}.
In our set--up we will also need a new ingredient, \ie\ 
the solution to (\ref{c-eq}) for the ghost field $c$,
which we will obtain  {\it ex novo}. 

However, a completely different path could  be followed:
indeed, we will show that it is possible to construct
the algebra {\it directly} on instanton moduli space, 
in particular without solving any field equation.
This is an important remark, since in this way the construction 
of the algebra  acquires a  
geometrical meaning and stands on its own.  
This approach is further developed in sec.\,\ref{poldo}, where we 
show its close relationship with the hyperk\"ahler quotient
construction of the instanton moduli space.
\subsection{Construction of the Solutions to the Equation of Motion
in the ADHM Formalism}

\label{derrick}
In sec.\,\ref{startrek} we saw that the TYM action localizes 
the fields relevant to the BRST transformations in (\ref{BRST}) onto 
a set of configurations dictated by a system of coupled differential 
equations. 
Here we will review  how to construct explicit solutions to these
equations of motion.

To begin with, recall that gauge fields  $A$ are projected 
onto  instanton configurations%
\footnote{In the previous section we adopted the standard convention 
in topological field theories of taking the gauge curvature 
to be anti--self--dual. Unfortunately
the literature on instanton calculus adopts 
the opposite convention (self--dual), to which 
we will conform from now on.}.
As it is well known, self--dual $SU(2)$ connections on $S^4$ can be 
put into one to one 
correspondence with holomorphic vector bundles of rank $2$ over  
$\C\P^3$
admitting a reduction of the structure group to its compact 
real form. The ADHM construction \cite{adhm} is an algorithm 
which gives all 
these holomorphic bundles and consequently all $SU(2)$ 
connections on $S^4$ (this $S^4$  should be thought of as  
the conformal compactification of $\Rq$. For the construction of 
instantons on $\Rq$, see for example \cite{dk}). 

The construction is purely algebraic and we find it more 
convenient to use quaternionic notations. The points, 
$x$, of the 
quaternionic space $\bb{H}\equiv \C^2\equiv\Rq$ 
can be conveniently
represented in the form $x=x^\mu \sigma_\mu$, with 
$\sigma_\mu=(i\sigma_c , \uno_{2\times 2}), c=1,2,3.$ 
The $\sigma_c$'s 
are the usual Pauli matrices, and
$\uno_{2\times 2}$ is the 2--dimensional identity matrix.
The conjugate of $x$ is $x^{\dagger} = 
x^\mu \bar{\sigma}_\mu$. 
A quaternion is said to be real if it is proportional to 
$\uno_{2\times 2}$ 
and imaginary if it has vanishing real part. 

The prescription to find an instanton of winding number $k$ is 
the following: introduce a $(k+1)\times k$ quaternionic matrix
linear in $x$
\beq
\Delta=a+bx \  ,
\label{f.4}
\eeq
where $a$ has the generic form 
\beq
\label{salute}
a=\pmatrix{w_1&\ldots&w_k\cr{}&{}&{}\cr{}&a'&{}\cr{}&{}&{}}\ \ ;
\eeq
$a'$ is a $k\times k$ quaternionic matrix.
The (anti--hermitean) gauge connection  is then written as 
\be
A=U^\dagger d U
\ \ ,
\label{trecinque}
\ee
where $U$ is a $(k+1)\times 1$ matrix of quaternions providing an
orthonormal frame of $\Ker \Delta^\dagger$, {\it i.e.}
\beqa
\Delta^\dagger U &=& 0
\ \ ,
\label{f.5}
\\
U^\dagger U &=&\uno_{2\times 2} 
\ \ .
\label{f.6}
\eeqa
The constraint (\ref{f.6}) ensures that $A$ 
is an element  of the Lie algebra of the $SU(2)$ gauge group.
The condition of self--duality 
\beq
\label{ddd}
{}^{\ast} F = F
\eeq
on the field strength of (\ref{trecinque}) 
is imposed by restricting the matrix $\Delta$ 
to obey
\be
\label{bos}
\Delta^\dagger\Delta=(\Delta^{\dagger}\Delta)^{T}
\ \ ,
\ee 
where the superscript $T$ stands for transposition of the
quaternionic
elements of the matrix (without transposing the quaternions 
themselves).
(\ref{bos}) in turn implies 
$\Delta^\dagger\Delta=f^{-1}\otimes\uno_{2\times 2}$, 
where $f$ is an invertible hermitean $k\times k$ matrix 
(of real numbers). 
From (\ref{trecinque}), the field strength of the gauge field
can be computed and it is
\be
F = U^\dagger d \Delta  f d \Delta ^\dagger U \ \ .
\label{bohboh2}
\ee
From this one can derive the following remarkable expressions for 
$\Tr (F F)$  \cite{corri,osbo} (see also \cite{maciocia})%
\footnote{The conventions used in this paper imply that 
the Pontryagin index is given by
$ -1/(8 \pi^2)\int   \Tr (F F)$, 
and it is positive (negative) on (anti--)instanton configurations.}: 
\beqa
\label{occam1}
\Tr (FF) 
&=& - {1\over 2} \Box \Box \ln {\rm det} \Delta^\dagger \Delta\ d^4 x
\\
\label{occam2}
&=& d\  \Tr \left[ P d D (d D)^\dagger D (d D)^\dagger
     + {1\over 3}  (D^\dagger d D)( D^\dagger dD)( D^\dagger dD )\right]     
\ \ ,
\eeqa
where 
\be
P=UU^\dagger=1-\Delta f\Delta^\dagger
\label{f.666}
\ee
is the projector on the kernel of 
$\Delta^\dagger$, and  according to \cite{corri}
the columns of $\Delta$, which are independent, have been
orthonormalized and collected into a matrix we have called $D$.

Gauge transformations are implemented in this formalism 
as  right multiplication   of $U$
by a unitary (possibly $x$--dependent) quaternion.
Moreover, $A$ is invariant  under 
reparametrizations of the ADHM data as follows:
\be
\label{nonbanane}
\Delta\rightarrow  Q\Delta R \ ,
\ee 
with $Q\in Sp(k+1), R\in GL(k,\real)$.  It is straightforward to see 
that (\ref{nonbanane}) preserves the bosonic constraint (\ref{bos}).
These symmetries  can be used to simplify the expressions
of $a$ and $b$. Exploiting this fact, in the
following we will 
choose the matrix $b$ to be
\be
b=-\pmatrix {0_{1\times k}\cr\uno_{k\times k}}.
\label{boh}
\ee
Choosing the canonical form (\ref{boh}) for $b$,
the bosonic constraint (\ref{bos}) becomes
\beqa
\label{a.1}
&&a^{\prime} = {a^\prime}^{T} \ \ ,
\\
&&a^\dagger a = (a^\dagger a  )^T \ \ .
\label{a.2}
\eeqa
Moreover, in this case there 
still exist  left--over $O(k)\times SU(2)$ reparametrizations 
of the form
(\ref{nonbanane}),  where now $R\in O(k)$,  
\beq\label{ahoo}
Q = \pmatrix{q&0&\ldots&0\cr
0&{}&{}&{}\cr \vdots&{}&R^T&{}\cr
0&{}&{}&{}}
\ \ , 
\eeq
and 
$q\in SU(2)$.
These transformations act non--trivially on 
the matrix $a$ and leave $b$  invariant. 
After imposing the constraint (\ref{bos}), the number of independent 
degrees of freedom contained in $\Delta$  
(that is the number of independent collective coordinates 
that the ADHM formalism uses to describe an instanton of winding number $k$)
is $8k + k(k-1) / 2$; modding  out  the $O(k)\times SU(2)$ 
reparametrization transformations, we would get $8k-3$ 
truly independent degrees of freedom. However (\ref{f.5}) and (\ref{f.6})
do not determine $U_0/ |U_0|$, where $U_0$ is the first component 
of $U$; this adds three extra degrees of freedom, so that in conclusion 
we end up with  a moduli space of dimension $8k$
(the instanton moduli space $\mathscr{M}^{+}$).
It is easy to convince oneself that the arbitrariness in $U_0/ |U_0|$
can be traded for the $SU(2)$ reparametrizations; in other words,
one can forget to mod out the $SU(2)$ factor of 
the reparametrization group $O(k)\times SU(2)$
but fix the phase of the quaternion $U_0$ (setting  for example
$U_0 = |U_0|\uno_{2\times 2}$). 
This is what we will actually do in the following.

We now focus our attention on the other fields involved in the BRST algebra
(\ref{BRST}). To begin with, the TYM action projects the anti--commuting
1--form $\psi_\mu$ onto the solutions to 
\be
{}^{\ast}( D_{[\mu} \psi_{\nu ]} ) = D_{[\mu} \psi_{\nu ]},\quad 
 D_{\mu} \psi_{\mu}=0 \ \ , 
\label{f.8}
\ee
where $D$ is the covariant derivative in the instanton 
background,  Eq.(\ref{trecinque}).
The solution to (\ref{f.8}) can be written as  \cite{osb}
\beq
\label{f.9}
\psi=U^{\dagger}{\cal M}f(d\Delta^{\dagger})U+
U^{\dagger}(d\Delta)f{\cal M}^{\dagger}U\ \ , 
\eeq
where ${\cal M}$ is a $(k+1)\times k$ matrix of quaternions, 
whose elements are Grassmann variables; moreover, in order 
for (\ref{f.9}) to satisfy (\ref{f.8}), ${\cal M}$  must obey the
constraint
\beq
\label{fconstr}
\Delta^{\dagger}{\cal M}=(\Delta^{\dagger}{\cal M})^T \ \ .
\eeq
(\ref{f.8}) tell us that the $\psi$ zero--modes are the tangent vectors 
to the instanton moduli space $\mathscr{M}^+$; as it is well known, 
the number of independent zero--modes  is $8k$ (the
dimension of $\mathscr{M}^+$), and we would like to see how 
this is implemented 
in the formalism of the ADHM construction. To this end, 
note that 
${\cal M}$ has 
$k(k+1)$ quaternionic elements ($4k(k+1)$ real degrees of freedom)
which are subject to the
$4k(k-1)$ constraints given by (\ref{fconstr}). The number of independent 
${\cal M}$'s 
satisfying (\ref{fconstr}) is thus $8k$,  as desired.

If we work in  the gauge in which $b$ has the canonical form
(\ref{boh}), then (\ref{fconstr}) can be conveniently elaborated as follows.
We put ${\cal M}$ in a form which parallels the one for $a$ in (\ref{salute}),
\ie\
\beq
\label{salutem}
{\cal M}=\pmatrix{\mu_1&\ldots&\mu_k\cr{}&{}&{}\cr{}&{\cal M}'&{}\cr{}&{}&{}}
\ \ , 
\eeq
${\cal M}^\prime$ being a $k\times k$ quaternionic matrix.
Plugging  (\ref{salute}), (\ref{boh}), (\ref{salutem})
into (\ref{fconstr}) we get
\beqa
\label{m.1}
&&{\cal M}^{\prime} = {{\cal M}^\prime}^{T} \ \ ,
\\
&&a^\dagger {\cal M} = (a^\dagger {\cal M}  )^T \ \ .
\label{m.2}
\eeqa
When $\Delta$ is transformed according to (\ref{nonbanane}),
the ${\cal M}$'s must also be reparametrized in such a way to keep 
the constraint (\ref{fconstr})  unchanged. This implies that 
the ${\cal M}$'s undergo the same formal reparametrization of
$\Delta$, that is 
\beq
\label{nonbananen}
{\cal M}
\rightarrow  Q{\cal M} R
\ \ .
\eeq
We now turn to the scalar field configuration. This is dictated by 
(\ref{phi-eq}), which should be supplemented by some boundary condition at 
infinity. Without loss of generality, we will set 
\be
\label{gsb}
\lim_{|x| \to \infty} \phi = {\cal A}_{00}=v\sigma^{3}/2i \ \ ,
\eeq
where $v\in \C$.
The solution to (\ref{phi-eq}) and (\ref{gsb}) 
was found in \cite{DKM} and reads 
\beq
\label{soltot}
\phi= U^{\dagger}{\cal M} f {\cal M}^{\dagger}U+U^{\dagger}{\cal A} U\ \ ,
\eeq
where 
\beq
\label{rott}
{\cal A}=\pmatrix{{\cal A}_{00}&0&\ldots&0\cr0&{}&{}&{}\cr
\vdots&{}&{\cal A}'&{}\cr 0&{}&{}&{}} \ \ .
\eeq
Here ${\cal A}'$
is a $k \times k$ real antisymmetric matrix, and 
the condition for (\ref{soltot}) to satisfy (\ref{phi-eq}) is
\beq
\label{eqnA}
\Delta^{\dagger}{\cal A}\Delta-(\Delta^{\dagger}{\cal
A}\Delta)^T=
- \Lambda_f
\ \ , 
\eeq
where we set 
\beq
\label{lambdaeffe}
\Lambda_f = {\cal M}^{\dagger}{\cal M} - ({\cal M}^{\dagger}{\cal M})^T 
\ \ ;
\eeq
note that $\Lambda_f = - \Lambda_{f}^{T} \propto \uno_{2\times 2}$.
Hereafter, by ${\cal A}\Delta$ we intend 
the $x$--independent $(k+1)\times k$ matrix 
\beq
\label{tothefacc}
{\cal A}\Delta=\pmatrix{{\cal A}_{00}w_1-w_m{\cal A}'_{m1}&
\ldots &{\cal A}_{00}w_k-w_m{\cal A}'_{mk}\cr\cr
{}&[{\cal A}',a']&{}\cr{}&{}&{}}
\ \ .
\eeq
(\ref{tothefacc})  is obtained by 
exploiting  the fact that,  in the calculations of the observables, 
expressions like ${\cal A}\Delta$ are always   
multiplied from
the left by $U^{\dagger}$; therefore,  recalling that 
$U^{\dagger}a = -U^{\dagger}bx$,  we can  eliminate the $x$--dependence
in ${\cal A}\Delta$. 
Using (\ref{salute}) and (\ref{tothefacc}), 
the $x$--dependence  
disappears  also from the l.h.s. of (\ref{eqnA}), 
which becomes
\beq
\label{cometichiami}
\Delta^{\dagger}{\cal A}\Delta-(\Delta^{\dagger}{\cal
A}\Delta)^T\equiv
L\cdot{\cal A}^{\prime} + \Lambda_{b} ({\cal A}_{00}) \ \ ;
\eeq
according to \cite{DKM},  the action of $L$ on
$k \times k$ matrices $\Omega^\prime$ is given by 
\beq 
\label{risc}
L\cdot\Omega^\prime
=-\frac{1}{2}\{\Omega^\prime,W\}+\frac{1}{2}\Tr 
\left([\bar{a}^{\prime},\Omega^\prime]a^{\prime}-\bar{a}^{\prime}[a^{\prime}
,\Omega^\prime]\right)\ \ ,
\eeq
where $W_{kl}=\bar{w}_{k}w_{l}+\bar{w}_{l}w_{k}$, 
and
\beq
\label{lb-def}
[\Lambda_{b}]_{ij} (\Omega_0) =  
\bar{w}_i \Omega_0 w_j - \bar{w}_j \Omega_0 w_i \ \ .
\eeq
Note that $[\Lambda_{b}]_{ij} (\Omega_0)$ are $c$--numbers when
$\Omega_0^\dagger = - \Omega_0$. 
(\ref{eqnA}) can be now  more compactly written as 
\beq
\label{eqnA2}
L\cdot{\cal A}^{\prime}= - \Lambda_{b} ({\cal A}_{00}) - \Lambda_f\ \ .
\eeq
The structure of (\ref{eqnA2}) suggests setting 
\beq
\label{decca}
{\cal A}^{\prime}= {\cal A}^{\prime}_b + {\cal A}^{\prime}_f 
\ \ ,
\eeq
where 
\beqa
\label{decca1}
L\cdot{\cal A}^{\prime}_b 
&=& - \Lambda_{b} ({\cal A}_{00}) \ \ ,
\\ 
\label{decca2}
L\cdot{\cal A}^{\prime}_f &=& - \Lambda_f
\ \ .
\eeqa 
This decomposition is useful since 
the solution $\phi_{\rm hom}$ to the homogeneous equation 
\beq
\label{fff}
D^{2}\phi =0
\eeq
with the non--trivial boundary condition  (\ref{gsb}) 
has the 
form 
\beq
\phi_{\rm hom}=U^{\dagger}{\cal A}_b U\ \ , 
\eeq
where we set 
\beq
\label{abos}
{\cal A}_b=\pmatrix{{\cal A}_{00}&0&\ldots&0\cr
0&{}&{}&{}\cr\vdots&{}&{\cal A}'_b&{}\cr 0&{}&{}&{}} 
\ \ ;
\eeq
using (\ref{cometichiami}), (\ref{decca1}) can be written as
\beq
\label{asshomeq}
\Delta^{\dagger}{\cal A}_b\Delta-(\Delta^{\dagger}{\cal
A}_b\Delta)^T = 0
\ \ ,
\eeq
which is the homogeneous equation associated to (\ref{eqnA}). 
Moreover, $L$ is a generally invertible operator 
acting on $k\times k$ matrices  \cite{DKM}. As a consequence,  
${\cal A}^{\prime}_b\neq 0$  if and only if 
${\cal A}_{00} \neq 0$. This is because non--trivial
solutions to the homogeneous equation (\ref{fff})
exist only 
when non--trivial boundary conditions on $\phi$ are imposed. 
On the other hand, the solution $\phi_{\rm inh}$ to 
(\ref{phi-eq}) supplemented by trivial boundary conditions
\be
\label{gsbb}
\lim_{|x| \to \infty} \phi_{\rm inh} = 0
\ \ ,
\ \ 
\eeq
reads as 
\beq
\phi_{\rm inh} = 
U^{\dagger}{\cal M} f {\cal M}^{\dagger}U+U^{\dagger}{\cal A}_f U\ \ ,
\eeq
with ${\cal A}_f$ given by 
\beq
\label{mah}
{\cal A}_f=\pmatrix{0&0&\ldots&0\cr
0&{}&{}&{}\cr\vdots&{}&{\cal A}'_f&{}\cr 0&{}&{}&{}} 
\ \ .
\eeq
As before, the reparametrization invariance (\ref{nonbanane})
induces a transformation on the matrix ${\cal A}$, which can be
found by requiring that the new  matrix 
still satisfies (\ref{eqnA}) when  $\Delta$ is replaced by  
its transformed expression; to this end one must have 
\beq
{\cal A}\rightarrow  Q {\cal A} Q^\dagger
\ \ .
\eeq
The last field relevant to our discussion is the ghost field $c$,
which in principle 
should be determined by solving (\ref{c-eq}). 
However, the definition for the universal connection $\widehat{A}$ given in 
(\ref{apiuci}), the expression  (\ref{dipiuesse}) for  $\widehat{d}$ 
together with the explicit form 
(\ref{trecinque}) for $A$ suggest a simple guess for its
ADHM expression; we write
\be
c=U^\dagger (s + {\cal C})  U 
\ \ , 
\label{trecin}
\ee
where ${\cal C}$
is the connection associated with the reparametrizations 
of the ADHM construction, which are shown in  (\ref{nonbanane}).
Therefore, under these symmetries  it transforms as
\be
\label{piove}
{\cal C}\rightarrow  Q ({\cal C} + s ) Q^\dagger
\ \ .
\ee
In sec.\,\ref{startrek} we observed that the first equation in (\ref{BRST}) 
together with (\ref{tmod}) imply that 
the BRST operator $s$ has an explicit realization 
on instanton moduli space as the exterior 
derivative. Since we are describing 
this space in terms
of a redundant parametrization, every expression should be covariant
with respect to the reparametrization symmetry group. 
This implies that ordinary derivatives
on the instanton moduli space (which is described by the 
redundant set of  $8k + k(k-1)/2$ 
ADHM collective coordinates) have to be replaced by 
covariant ones, and $s$ by 
its covariant counterpart
\beq
\label{dieci}
{\cal S} = s + {\cal C} \ \ ,
\eeq
which is exactly what appears in (\ref{trecin}). 
The criterion to fix ${\cal C}$ is clear: one has simply to plug 
(\ref{trecin}) into (\ref{c-eq}),
and solve for ${\cal C}$. 
In the next section we will illustrate
an alternative (and quicker) way to construct ${\cal C}$.
At this point we must alert the reader that the situation in which
${\cal A}_{00}\neq 0$ 
requires a more detailed analysis. 
As we will discuss in subsec.\,\ref{spazio1999},  in  this
case the correct guess for  ${\cal C}$ is
\beq
\label{alert}
{\cal C}=\pmatrix{{\cal C}_{00}&0&\ldots&0\cr0&{}&{}&{}\cr
\vdots&{}&{\cal C}'&{}\cr 0&{}&{}&{}} \ \ ,
\eeq
where ${\cal C}'= - ({\cal C}')^T$ and 
${\cal C}_{00}$ is non--vanishing when ${\cal A}_{00}\neq 0$.
For the moment, let us only observe that (at least when ${\cal A}_{00}= 0$),
${\cal C}$ is from its very definition 
a moduli--dependent $(k+1)\times (k+1)$ matrix 
antisymmetric in its lowest $k\times k$ block and zero elsewhere.

In summary, we are left with four ADHM matrices:
\begin{itemize}
\item[\bf 1.]
$\Delta$, which collects the ADHM data of the instanton configuration
($8k + k(k-1) / 2$ degrees of freedom), 
\item[\bf 2.]
${\cal M}$, which parametrizes the tangent vectors to the instanton 
moduli space ($8k$ degrees of freedom), 
\item[\bf 3.]
${\cal A}$, which is the solution to (\ref{eqnA}), and
\item[\bf 4.]
${\cal C}$, which is the matrix in (\ref{alert}).
\end{itemize}
Under the action of the group of reparametrization of the ADHM construction,
$\Delta $ and ${\cal M}$ transform in the fundamental representation,
whereas ${\cal C}$ transforms as a connection and 
${\cal A}$ as the curvature of a connection. 
We want to warn the reader that only 
$\Delta$ and ${\cal C}$ will emerge as independent quantities. 
Once they are given, all the other quantities  
(\ie\ ${\cal M}$ and ${\cal A}$) will be completely determined,
as we will show in the next subsection. 

\subsection{The BRST Algebra in the ADHM Formalism: the Zero Vacuum
Expectation Value Case}
\label{wepper}

As we have seen, 
the TYM  action projects the field  $A$
onto the solutions of the self--duality equations (\ref{ddd}), and 
the anti--commuting 1--form $\psi$  onto the tangent vectors to the
instanton moduli space (the solutions to (\ref{f.8})); moreover, the
scalar field $\phi$ must satisfy 
(\ref{phi-eq}), possibly with its boundary condition (\ref{gsb}),
and the ghost field $c$ satisfies  (\ref{c-eq}), which is induced by the 
transversality condition of $\psi$ in the instanton background.  
In the last section we used the ADHM formalism to 
write the solutions to these coupled equations,
that we collect here for the sake of clarity:
\beqa 
\label{brstfields}
A&=&U^{\dagger}dU\ \ ,
\nonumber\\ 
c&=&U^{\dagger}(s+{\cal C})U\ \ ,
\nonumber\\  
\psi&=&U^{\dagger}{\cal M}f(d\Delta)^{\dagger}U+
U^{\dagger}(d\Delta)f{\cal M}^{\dagger}U\ \ ,
\nonumber\\
\phi&=&U^{\dagger}{\cal M}f{\cal M}^{\dagger}U+U^{\dagger}{\cal A}U
\ \ .
\eeqa 
The BRST transformations of these  fields are written in (\ref{BRST}).
If we now plug (\ref{brstfields}) into (\ref{BRST}), we end with a set of 
equations which will provide us 
with explicit expressions for 
the variations $s\Delta$, $s {\cal M}$, $s {\cal C}$, $s {\cal A}$ in terms
of $\Delta$, ${\cal M}$, ${\cal C}$, ${\cal A}$.
At the same time we will also show how to determine the explicit 
form of $ {\cal C}$.

A preliminary  ingredient which is necessary for this 
computation is the knowledge of
$s U$; we would like to express it in terms of $s \Delta$, otherwise 
we would be forced to 
solve the highly non--trivial set of algebraic 
equations (\ref{f.5}), (\ref{f.6}) for $U$.
The following trick is then useful. Perform the BRST variation of 
(\ref{f.5}), 
\beq
(s\Delta )^{\dagger}  U + \Delta^{\dagger} s U = 0 \ \ ;
\eeq
this can be read as an equation for $s\Delta$, whose solution is%
\footnote{The following expression for $sU$ would still be valid 
if $s$ would represent  a generic variation.}  
\beq
s U = - \Delta f (s \Delta )^{\dagger} U + U (U^\dagger s U)
\ \ . 
\eeq
We are now in a position to start computing $sA$ by varying the 
first  equation of 
(\ref{brstfields}). 
This way we get 
\be
\label{nuvole2}
s A = U^{\dagger} \Bigl[ s \Delta f (d \Delta)^{\dagger} + d \Delta f (s 
\Delta)^{\dagger} \Bigr] U - [D, U^{\dagger} s U ]
\ \ ,
\ee
where, for a generic 1--form $K$, we put $[D, K]= d K + AK + KA$.
Here and in the following we repeatedly use the fact that 
\be
\Delta^{\dagger} d U = -(d \Delta)^{\dagger} U
\ \ ,
\ee
which is a consequence of (\ref{f.5}).
We now substitute the explicit expressions found for $\psi$ and $c$
into the r.h.s. of the first of (\ref{BRST}), thus getting 
\beqa
\label{nuvole}
\psi-Dc &=&
U^{\dagger} ( {\cal M} f d \Delta^{\dagger} + d \Delta f 
{\cal M}^{\dagger} )U  + \nonumber\\
&&
- [D, U^{\dagger} s U ]  - [D, U^{\dagger} {\cal C} U ]
\ \ .
\eeqa
If we equate the r.h.s. of (\ref{nuvole}) to the r.h.s. of 
(\ref{nuvole2}) we obtain, after a little algebra,
\beq
U^{\dagger} ( {\cal M} f d \Delta^{\dagger} + d \Delta f 
{\cal M}^{\dagger} )U  = 
U^{\dagger} \Bigl[ (s \Delta + {\cal C} \Delta) f d \Delta^{\dagger} + 
d \Delta f  (s \Delta + {\cal C} \Delta)^{\dagger} \Bigr] U \ \ ;
\eeq
from here we conclude that 
\be
\label{eccheneso}
{\cal M} = s \Delta + {\cal C} \Delta
\ee
modulo ``irrelevant" terms, that is terms which vanish when right (left)
multiplied by $U$ ($U^{\dagger}$). 
The same strategy as before can be repeatedly applied to the remaining
equations in (\ref{BRST}), thus obtaining 
the complete action of the BRST operator on  
$\Delta$, ${\cal M}$, ${\cal C}$, ${\cal A}$. The result of this exercise is
\beqa
\label{azzarolina}
s\Delta&=&{\cal M}-{\cal C}\Delta\ \ ,\nonumber\\
s{\cal M}&=&{\cal A}\Delta-{\cal C}{\cal M}\ \ ,\nonumber\\
s{\cal A}&=&-[{\cal C},{\cal A}]\ \ ,\\
s{\cal C}&=&{\cal A}-{\cal C} {\cal C}\ \ ,\nonumber
\eeqa
which is the realization of the BRST algebra on the instanton moduli
space.%
\footnote{Using (\ref{tothefacc}) and  similar expressions for 
${\cal A}{\cal M}$, ${\cal C}\Delta$ , ${\cal C}{\cal M}$, 
the $x$--dependence completely 
disappears from (\ref{azzarolina}).}

Three observations are in order. First, it is 
straightforward to show that $s^2$ is  nilpotent as it should.
This can be simply done by applying once again $s$ 
to each equation in (\ref{azzarolina}).  
Therefore, {\it on instanton moduli space $s$ is the exterior
derivative}, as we announced in the previous sections.   
Second, the last two equations in  (\ref{azzarolina})
and the nilpotency of $s$ suggest that 
${\cal A}$ can be interpreted as the curvature of the connection 
${\cal C}$ (these equations then becoming the Bianchi identity for 
${\cal A}$ and its definition in terms of ${\cal C}$).
Last, 
using the covariant derivative defined in (\ref{dieci}),
we can rewrite the BRST algebra on  instanton moduli space in
a more compact form as  
\beqa
\label{azzarolinabisse}
{\cal S} \Delta&=&{\cal M}\ \ ,\nonumber\\
{\cal S} {\cal M}&=&{\cal A}\Delta \ \ ,\nonumber\\
{\cal S}{\cal A}&=& 0\ \ ,\\
s\,{\cal C}+{\cal C}{\cal C}&=&{\cal A} \ \ .\nonumber
\eeqa

We now discuss the important point of how to compute 
the connection ${\cal C}$. 
This can be done by plugging the first equation
of (\ref{azzarolinabisse}) into the fermionic constraint
(\ref{fconstr}), thus getting 
\beq
\Delta^{\dagger}{\cal C}\Delta-
\bigl(\Delta^{\dagger}{\cal C}\Delta\bigr)^T
=(\Delta^{\dagger}s\Delta)^T-\Delta^{\dagger}s\Delta\ \ .
\label{C-eq}
\eeq
It can also be shown  that this equation is equivalent to 
the ADHM transcription of (\ref{c-eq}). 
In the following considerations 
we restrict  our attention to the case 
in which ${\cal C}_{00} = 0$ (recall (\ref{alert})); as we said in the last 
section, this is true if and only 
if ${\cal A}_{00} = 0$. The case 
$({\cal C}_{00}, {\cal A}_{00}) \neq (0,0)$ is crucially different and 
deeply related to the fact that when the scalar field 
acquires a non--zero v.e.v., the theory has a new invariance
(the $U(1)$ central charge  symmetry). For these reasons
it will be separately analysed in 
sec.\ref{spazio1999}.
In this section we limit ourselves to ${\cal C}$'s of the form 
\beq
\label{alert2}
{\cal C}_f=\pmatrix{0&0&\ldots&0\cr0&{}&{}&{}\cr
\vdots&{}&{\cal C}'_{f}&{}\cr 0&{}&{}&{}} \ \ ,
\eeq
where ${\cal C}^\prime_f= - ({\cal C}^\prime_f)^T$.
Using the expression for the operator $L$ introduced in 
(\ref{risc}), we can rewrite  (\ref{C-eq}) more compactly as 
\beq
L\cdot{\cal C}^\prime_f = -  \L_{{\cal C}} \ \ , 
\label{cipreq}
\eeq
where we have defined
\beq
\L_{{\cal C}} \equiv \Delta^{\dagger}s\Delta -(\Delta^{\dagger}s\Delta)^T 
\ \ .
\label{LC}
\eeq
The solution to (\ref{cipreq}) is then formally written as 
\beq
{\cal C}^\prime_f = - L^{-1}\cdot \L_{{\cal C}} \ \ ,
\label{Cf}
\eeq
due to the invertibility of $L$.
 
The r\^{o}le of the connection ${\cal C}$ can be conveniently elucidated 
by  setting it to zero in (\ref{brstfields}); in this case the BRST algebra 
would read%
\footnote{The following transformations
are in close relationship to the supersymmetry transformations of the ADHM 
matrices given in \cite{DKM}.} 
\beqa\label{mizzica}
s\Delta&=&{\cal M}\ \ ,\nonumber\\
s{\cal M}&=&{\cal A}\Delta\ \ ,\\
s{\cal A}&=&0\ \ .
\nonumber
\eeqa
It can be immediately shown that in this case the operator $s$ would 
fail to be nilpotent. Indeed, the action of 
$s^2$ on the ADHM matrices would become 
\beqa\label{gcr3}
s^2\Delta&=&{\cal A}\Delta\ \ ,\nonumber\\
s^2{\cal M}&=&{\cal A}{\cal M}\ \ , \\
s^2{\cal A}&=&0\ \ .\nonumber
\eeqa
$s^2$ would then be nilpotent only up to 
transformations generated by $k\times k$ {\it moduli-dependent} 
antisymmetric matrices, {\it i.e.} local  
reparametrizations in the moduli space.
(\ref{gcr3}) are the transcription of (\ref{delta2}) on the moduli
space. 
 
In summary, the universal connection $\widehat A$ is given by
\beq
\widehat A = U^{\dagger}(d+s+{\cal C})U \ \ .
\label{apiucf}
\eeq
We want now to comment on the interpretation of the results 
obtained in this section.
The crucial observation is that,  
once (\ref{apiucf}) is given, the ADHM matrices
${\cal M}$ and ${\cal A}$ are in turn determined by (\ref{azzarolina}) 
as the covariant derivative of $\Delta$ and the
curvature of the connection ${\cal C}$ respectively;
the only independent variables are the collective coordinates contained in 
$\Delta$ (the instanton moduli and other moduli possibly associated with
redundancies of the ADHM parametrization) and their differentials
(the entries of the matrix $s\Delta$). 
Once the reparametrization invariance has been  gauged away 
(by giving some convenient prescription; see the explicit 
examples in sec.\,\ref{victor}), physical quantities  
become, through their ADHM expression,   differential forms%
\footnote{When scalar fields have non--zero v.e.v.'s this picture is 
slightly modified; we postpone this discussion to sec.\,\ref{spazio1999}.} 
on 
the $8k$--dimensional (anti--)instanton moduli space $\mathscr{M}^{+}$
($\mathscr{M}^{-}$).
Operatively, this amounts to first 
identifying a correct parametrization  for the instanton configuration 
(in term of the ADHM matrix $\Delta$ introduced in ({\ref{bos})), 
and then to computing  the explicit expression
for the 1--form ${\cal C}$ using  (\ref{fconstr}), in which
${\cal M}$ is substituted by its expression (\ref{eccheneso}).
Finally,  ${\cal A}$ is 
determined by the last equation in (\ref{azzarolina}).

\subsection{The BRST Algebra in the ADHM Formalism: the Non--Zero 
Vacuum Expectation Value Case}
\label{spazio1999}
The realization of the BRST algebra (\ref{brstZ}) on instanton moduli
space in the case in which scalar fields have non--vanishing v.e.v.
closely parallels that of sec.\,\ref{wepper}. 
In particular, the universal connection
$\widehat{A} =  A+c+\L$ is again expressed as 
\beq
\widehat{A} = U^\dagger (d + s + {\cal C}) U \ \ ,
\label{a-hat}
\eeq
and its curvature equation and Bianchi identities give rise to the 
same algebra (\ref{azzarolina}) for the ADHM matrices.
Notice however that 
in this case ${\cal C}$ is given by
\beq
{\cal C}=\pmatrix{{\cal C}_{00}&0&\cdots&0 \cr 0&{}&{}&{}\cr
\vdots&{}&{\cal C}^{\prime}&{}\cr 0&{}&{}&{}}\ \ ,
\label{C}
\eeq
the element ${\cal C}_{00}$ of (\ref{C}) being related to the asymptotic
behavior of the ghost $c+\L$ at $|x| \rightarrow \infty$, 
\beqa
\lim_{|x| \rightarrow \infty}(c + \L) \equiv 
\lim_{|x| \rightarrow \infty} U^\dagger (s + {\cal C}) U = {\cal C}_{00} \ \ .
\label{C-infty}
\eeqa
From (\ref{L-bc}) it then follows 
\beq
{\cal C}_{00} = \lambda  {\cal A}_{00} \ \ .
\label{C00}
\eeq
The ADHM connection ${\cal C}$ can be calculated by solving
(\ref{C-eq}); since  
\beq
\Delta^{\dagger}{\cal C}\Delta-(\Delta^{\dagger}{\cal
C}\Delta)^T \equiv
L\cdot{\cal C}^{\prime} + \Lambda_b ({\cal C}_{00})\ \ ,
\label{neweqnA}
\eeq
then 
(\ref{C-eq}) can be written as 
\beq
\label{C-eq2}
L\cdot{\cal C}^{\prime}  =
- \Lambda_b ( {\cal C}_{00})- \Lambda_{{\cal C}}  \ \ ,
\eeq
where  $\Lambda_b$ has been defined in (\ref{lb-def}), and
$\L_{{\cal C}}$ is given by (\ref{LC}).
(\ref{C-eq2}) and  (\ref{C-eq}) are formally identical to (\ref{eqnA2}}), 
(\ref{eqnA}) respectively; 
as in sec.\,\ref{derrick}, they suggest us to set 
\beq
\label{decc}
{\cal C} = {\cal C}_b + {\cal C}_f 
\ \ , 
\eeq
where 
${\cal C}_b$ satisfies  the associated homogeneous equation 
\beq
\label{eqnCbos}
\Delta^{\dagger}{\cal C}_b\Delta-(\Delta^{\dagger}{\cal
C}_b\Delta)^T=0\ \ .
\eeq
The solution to (\ref{eqnCbos}) is (unique and) 
completely specified only after imposing 
boundary conditions, as in (\ref{C00}).
This reflects, in the ADHM language,
the degeneracy (\ref{tmod-Z}) in the definition of tangent vector to
the instanton moduli space, which is due to the existence of 
central charge transformations; we know in fact that the
fermionic constraint (\ref{fconstr}), from which (\ref{C-eq}) 
directly follows, 
is just the ADHM transcription of the fermionic zero--mode equations 
(\ref{f.8}).
As in that case, the solution is unique once the non--trivial boundary 
condition (\ref{C00}) is imposed.%
\footnote{See also the discussion after (\ref{cipreqbis}).} 
Let us now set 
\beq
{\cal C}^{\prime} ={\cal C}^{\prime}_f + {\cal C}^{\prime}_b
\ \ ;
\eeq
then (\ref{C-eq2}) gives 
\beqa
L \cdot {\cal C}^{\prime}_b &=& - \Lambda_b ( {\cal C}_{00})
\ \ ,
\\ \nonumber 
L\cdot{\cal C}^\prime_f &=& -  \L_{{\cal C}} \ \ , 
\eeqa
whose solution is unique once ${\cal C}_{00}$ has been specified 
by means of the boundary condition (\ref{C00}).
Note that, if ${\cal C}_{00}$ were zero, then also 
$\Lambda_b ({\cal C}_{00})$ would vanish; therefore, 
due to the invertibility of $L$, the
equation for ${\cal C}^{\prime}_b$ would only admit the trivial solution
${\cal C}^{\prime}_b=0$.

The matrices ${\cal M}$ and ${\cal A}$ are in turn determined by means
of the ADHM algebra to be 
\beqa
&& {\cal M} = {\cal S} \Delta \ \ , \nonumber\\
&& {\cal A} = s{\cal C} + {\cal C}{\cal C} \ \ ;
\eeqa
in particular, for the $(00)$ element of ${\cal A}$, we have
\beqa
&& {\cal A}_{00} = s {\cal C}_{00} = {\partial\over{\partial\lambda}}
\Bigl( \lambda  v \frac{\sigma_{3}}{2i}\Bigr) =  v \frac{\sigma_{3}}{2i}
 \ \ , \nonumber\\
&& s {\cal A}_{00} = {\partial\over{\partial\lambda}} 
\Bigl( v \frac{\sigma_{3}}{2i}\Bigr) = 0 \ \ ,
\label{A00} 
\eeqa
from which it follows the expected asymptotic behavior (\ref{phi-bc}) 
for the scalar field $\phi$
\beq
\lim_{|x| \rightarrow \infty}\phi \equiv 
\lim_{|x| \rightarrow \infty} U^\dagger{\cal A} U = 
{\cal A}_{00} =  v \frac{\sigma_{3}}{2i} \ \ .
\eeq
Note that the nilpotent BRST operator $s$ acts on the external
parameters ${\cal C}_{00}, {\cal A}_{00}$, given respectively 
by (\ref{C00}) and (\ref{A00}), just as the partial derivative 
with respect to $\lambda$, 
while its restriction on the other elements of the 
ADHM matrices would act as the usual
exterior derivative on the moduli space. 

\subsection{Algebraic Construction of the BRST Transformations}
\label{algebraic} 
In this section we will  derive the realization of the BRST algebra 
on the instanton moduli space in a direct way, \ie\ 
using  neither the BRST algebra in field space (\ref{BRST}) 
nor the expressions  for the field configuration  (\ref{brstfields}) 
onto which the TYM action projects.
The only  ingredient we  need will be a 
parametrization for the moduli space of instantons; 
in terms of the ADHM construction,  this is equivalent to 
determine a matrix $\Delta$ which satisfies (\ref{bos}). 
As discussed in sec.\,\ref{derrick}, 
the ADHM space of parameters  is acted upon by 
an $O(k)$  reparametrization symmetry.
The gauging of this symmetry will turn out to be what is required 
to make the BRST variations of the ADHM data $\Delta$ consistent 
with the algebraic constraints (\ref{bos}) which determine them. 
The BRST algebra on instanton moduli space, (\ref{azzarolina}),
will thus emerge as the most general 
set of deformations of the ADHM data compatible
with  (\ref{bos}). 

To show this, let us now start by performing an infinitesimal  
scalar variation (that we call $s$ for obvious reasons)
of the bosonic constraint (\ref{bos}). We get 
\beq
\label{sbos}
(s\Delta)^{\dagger}\Delta+\Delta^{\dagger}s\Delta=[(s\Delta)^{\dagger}
\Delta]^T+(\Delta^{\dagger}s\Delta)^T\ \ .
\eeq
This relation should be read as an equation for $s \Delta$, and 
we want to guess its solution. We write it as 
\beq
\label{sdelta}
s\Delta={\cal M}-{\cal C}\Delta\ \ ,
\eeq
where ${\cal M}$ is {\it defined} as the matrix which satisfies 
(\ref{fconstr}).
${\cal C}$ is constrained by the
structure of $\Delta$  (which 
satisfies  (\ref{a.1}), (\ref{a.2}) in the gauge defined by (\ref{boh}))
and ${\cal M}$ (which is is fixed
by (\ref{m.1}), (\ref{m.2}));
in conclusion,  the most general expression  of  ${\cal C}$ consistent with 
(\ref{sdelta}) is 
\beq 
{\cal C}=\pmatrix{{\cal C}_{00}&0&\ldots&0\cr 0&{}&{}&{}\cr 
\vdots&{}&{\cal C}'&{}\cr
0&{}&{}&{}} \ \ ,
\eeq
where ${\cal C}'$ is a real antisymmetric $k\times k $ matrix, 
$({\cal C}^{\prime})^{\dagger}
=-{\cal C}^{\prime}$.
If we plug (\ref{sdelta}) into (\ref{sbos}), 
the terms containing ${\cal M}$ exactly cancel out 
thanks to  (\ref{fconstr}), whereas ${\cal C}$ is fixed by
the equation  
\beq
\Delta^\dagger ({\cal C} + {\cal C}^\dagger) \Delta = 
\Big[ \Delta^\dagger ({\cal C} + {\cal C}^\dagger) \Delta \Bigr]^T
\ \ , 
\eeq
which becomes
\beq
L\cdot ( {\cal C}^\prime + {{\cal C}^\prime}^\dagger )= 
-  \L_b ({\cal C}_{00} + {\cal C}_{00}^{\dagger})
\ \ ,
\label{cipreqbis}
\eeq
where $L$ is a generally invertible operator. 
As a consequence one must have ${\cal C}_{00} = - {\cal C}_{00}^{\dagger}$. 
One is thus led to an expression of ${\cal C}$ which coincides with 
the one previously suggested in (\ref{alert}). 

Let us now pause for a moment and count the number of degrees of freedom
in (\ref{sdelta}).
On one hand, the ADHM matrix 
$\Delta$ and its variation $s\Delta$ both contain   
$8k+k(k-1)/2$ (unconstrained)  degrees of freedom%
\footnote{We recall the reader that 
we always work with a canonical choice of $b$.}; 
${\cal M}$ contains  instead $8k$ degrees of freedom, after 
solving  (\ref{fconstr}). On the  other hand,  
${\cal C}$ contains $3 + k(k-1)/2$ parameters, and we would be led to an 
apparent mismatch in counting the number of degrees 
of freedom in (\ref{sdelta}). 
Actually 
the three degrees of freedom introduced by ${\cal C}_{00}$ are not
new; instead they
are already included in the number of independent solutions
to (\ref{fconstr}). This can be understood once we 
decompose ${\cal C}$ as in (\ref{decc}), where 
\beq 
{\cal C}_b=\pmatrix{{\cal C}_{00}&0&\ldots&0\cr 0&{}&{}&{}\cr 
\vdots&{}&{\cal C}^{\prime}_{b}&{}\cr
0&{}&{}&{}} \ \ 
\eeq
is defined in such a way to satisfy the homogeneous equation (\ref{eqnCbos}).  
On one hand this equation is equivalent to 
\beq
L\cdot{\cal C}^{\prime}_{b} =  - \Lambda_{b} ({\cal C}_{00}) 
\ \ ;
\eeq
therefore  it has non--trivial solutions when ${\cal C}_{00}\neq 0$.
On the other hand,  it is formally identical to (\ref{fconstr}) with
 ${\cal M}$ replaced by ${\cal C}_b \Delta$. 
In order to avoid double counting, 
its three independent solutions 
should then not be considered as ``new".
Finally, ${\cal C}_f$ is now constrained by (\ref{decc}), (\ref{sdelta}),
(\ref{fconstr}) to satisfy an equation identical to (\ref{C-eq}); 
thus,  it  just takes into account the genuinely new 
$k(k-1)/2$ parameters which are related to the $O(k)$ reparametrization 
invariance. We then conclude that there is 
a complete balance in the number of degrees of freedom in  (\ref{sdelta}). 
 
If we perform  the $s$--variation of the fermionic constraint
(\ref{fconstr}), we get  
\beq
(s\Delta)^{\dagger}{\cal M}+\Delta^{\dagger}s{\cal M}=
[(s\Delta)^{\dagger}{\cal M}]^T+(\Delta^{\dagger}s{\cal M})^T\ \ .
\eeq
Taking into account (\ref{sdelta}), we find
\beq
\label{sferm}
({\cal M}^{\dagger}+\Delta^{\dagger}{\cal C}){\cal M}-
[({\cal M}^{\dagger}+\Delta^{\dagger}{\cal C}){\cal M}]^T=
(\Delta^{\dagger}s{\cal M})^T-\Delta^{\dagger}s{\cal M}\ \ ; 
\eeq
(\ref{sferm}) should be thought of as an equation for 
$s{\cal M}$. Analogously to the previous case, 
its  most general solution can be cast into the form 
\beq
\label{meme}
s{\cal M}={\cal A}\Delta-{\cal C}{\cal M}\ \ ,
\eeq
where ${\cal A}$ has the same form as in (\ref{rott}).
If we plug (\ref{meme}) into (\ref{sferm}), we obtain that 
${\cal A}$ must satisfy the following relation:
\beq
\label{eqnA3}
\Delta^{\dagger}{\cal A}\Delta-(\Delta^{\dagger}{\cal
A}\Delta)^T=({\cal M}^{\dagger}{\cal M})^T-{\cal M}^{\dagger}{\cal
M}\ \ . 
\eeq
Then (\ref{eqnA3}) is identical to (\ref{eqnA}), 
which was obtained from a completely different point of view
(the equations of motion for the scalar field $\phi$),  and
its solution is given by 
(\ref{decca}), (\ref{decca1}), (\ref{decca2}).

We want now to clarify   the relation between 
${\cal A}$ and ${\cal C}$ as defined in this section. 
To this end, let us perform one more $s$--variation of 
(\ref{sdelta}) and  (\ref{meme});
after a little algebra we get
\beqa
s^2 \Delta &=& \Bigl( {\cal A}- s{\cal C} - {\cal C} {\cal C}\Bigr)\Delta 
\ \ ,
\nonumber\\
s^2 {\cal M}&=&  \Bigl( {\cal A}- s{\cal C} - {\cal C} {\cal C}\Bigr) {\cal M}
+ \Bigl( s {\cal A} + [ {\cal C}, {\cal A}]\Bigr) \Delta 
\ \ .
\eeqa
Once one requires the nilpotency of  the BRST operator $s$,  
then 
\beqa
\label{sera1}
&& {\cal A}- s{\cal C} - {\cal C}{\cal C} = 0
\ \ ,
\\
\label{sera2}
&&s {\cal A} + [ {\cal C}, {\cal A}] = 0
\ \ .
\eeqa
Therefore it is possible 
to interpret (\ref{sera1}) as the definition of ${\cal A}$ as the field 
strength of ${\cal C}$
and (\ref{sera2}) as its Bianchi identity. 
This completely clarifies  the relation between 
${\cal A}$ and ${\cal C}$. 

In order to check the consistency of the super--constraints
with the BRST variations, we still have to perform the $s$--variation of 
(\ref{eqnA3}).
If we do this, we get
\beq
\Delta^{\dagger}({s\cal A}+[{\cal C},{\cal A}])\Delta - [\Delta^{\dagger}
(s{\cal A}+[{\cal C},{\cal A}])\Delta]^T=0 \ \ ,
\eeq
which is trivially satisfied thanks to  (\ref{sera2}).

Summarizing, we have found that consistency between  the 
BRST variation   of the bosonic ADHM matrix $\Delta$ and the 
constraint (\ref{bos}) it obeys, yields  
\beqa
s\Delta&=&{\cal M}-{\cal C} \Delta\ \ ,
\nonumber\\
s{\cal M}&=&{\cal A} \Delta-{\cal C}{\cal M}\ \ ,
\nonumber\\
s{\cal A}&=&-[{\cal C},{\cal A}]\ \ ,
\\
s{\cal C}&=&{\cal A}-{\cal C} {\cal C}\ \ ,
\nonumber
\eeqa
where ${\cal M}$ satisfies (\ref{fconstr}). 
As anticipated, this set of equations gives an  explicit 
realization of the BRST algebra on instanton moduli space, and 
it coincides with that  
found in  (\ref{azzarolina}) with completely different methods.


\section{The Set--Up of the Calculation of Instanton Green's Functions}
\setcounter{equation}{0}
\label{ilaria}
In this section we explain how to perform instanton calculations
in our picture. As an application of our techniques, 
we will then focus on computing correlators in 
the case in which the relevant instanton 
configurations have winding   number $k=1,2$. 
In $N=2$ SYM with non--vanishing v.e.v. for the scalar field, 
we will be interested in evaluating the correlator $<\Tr\phi^2>$. 
These computations will show the main features of the formalism developed
in the previous sections.

To make these characteristics more evident,
let us now summarize the ``standard'' strategy to perform instanton 
calculations  in SUSY theories \cite{nsvz,akmrv,DKM,ft}.
\begin{itemize}
\item[{\bf 1.}] 
The action is expanded around the saddle point up to 
quadratic fluctuations.
\item[{\bf 2.}]  
The fields are expanded in eigenmodes and the functional measure is
replaced by an integration over the coefficients of the mode expansion.
The contribution of the zero--modes and that of the non--zero modes are
now clearly identified. 
\item[{\bf 3.}]  
The fields in the correlator are also expanded in modes and the
 part containing the non--zero modes is discarded since it
 represents higher order quantum corrections.
\item[{\bf 4.}]  
The non--zero modes are then integrated out. This integration
 gives a ratio of determinants which is one thanks to SUSY
 \cite{divecchia}. 
\item[{\bf 5.}]  
The last step consists in performing 
the integration over the zero--modes.
In order to deal with the zero--mode sector, 
one has to trade integrations over the bosonic 
zero--modes for integrations over collective coordinates; 
this gives rise to a bosonic Jacobian.
Moreover, one has to keep 
into account chiral selection rules which single out
the non--vanishing Green's functions.
Operatively, these selection rules amount
to say that all the Grassmann integrations over the fermionic
collective coordinates have to be 
soaked up by explicitly inserting the appropriate number (say $n$) of
zero--modes; 
thus, the only non--zero amplitudes will be those which admit an 
expansion in terms of fermion zero--modes such that the coefficient
multiplying the term with $n$ fermionic collective coordinates 
does not vanish. 
This gives rise to a fermionic Jacobian,
which is the determinant of the matrix whose entries are the
overlaps of the fermionic zero--mode wave functions. 
\end{itemize}

Our starting point will be the last step which, in
the formalism of the previous sections, amounts to integrating the
Lagrange multipliers in the gauge fixed TYM action (\ref{ztym-act}). 
This integration naturally projects the fields 
$A$, $\psi$, $\phi$
onto the zero--modes
subspace, which is identified by 
(\ref{ddd}), (\ref{f.8}) and (\ref{phi-eq})
(supplemented by appropriate boundary conditions on $\phi$)%
\footnote{If one wanted  to work with anti--instantons,
then (\ref{ddd}) and  (\ref{f.8})
should be replaced by (\ref{mod}) and  (\ref{tmod}), respectively.};
the configurations which solve these equations were written in 
(\ref{brstfields}).
Through these expressions,  physical amplitudes will depend on $\Delta$,
${\cal C}$, ${\cal M}$ and ${\cal A}$.
The ADHM equations (\ref{bos}) fix the number of independent 
(bosonic) collective 
coordinates to be $8k + k(k-1)/2$; 
gauge--fixing the left--over $O(k)$ symmetry 
further reduce this number to $8k$. 
Moreover the first relation in (\ref{azzarolina}) together with 
(\ref{fconstr}) allows one to  compute the connection 
${\cal C}^\prime$ as a 1--form 
expanded on a basis of differentials of the bosonic moduli. 
If one substitutes back the computed expression for ${\cal C}$ into 
the first equation in (\ref{azzarolina}), then the ${\cal M}$'s
become in turn differential 1--forms 
on instanton moduli space $\mathscr{M}^+$.%
\footnote{Recall that the 
the number of independent ${\cal M}$'s 
is $8k$ (as the number of bosonic moduli) by virtue of (\ref{fconstr}).} 
Finally (\ref{eqnA}) gives ${\cal A}$ as a function of 
$\Delta$ and ${\cal M}$. 
We then conclude that 
{\it any polynomial in the fields becomes, after projection
onto the zero--mode subspace, a well--defined  
differential form on} $\mathscr{M}^+$ \cite{witten}.
We  can then symbolically  write
\beq
\label{prescription}
\left< fields \right> = \int_{\mathscr{M}^{+}}  \ 
\left[ (fields)\ e^{-S_{\rm TYM}} \right]_{zero-mode\  subspace}
\ \ .
\eeq

Let us now call $\{ \widehat{\Delta}_{i}\}$ ($\{\widehat{\cal M}_i\}$), 
$i=1,\ldots, p$, where $p=8k$, a basis of (ADHM) coordinates on 
$\mathscr{M}^+$ ($T_A\mathscr{M}^+$).
(\ref{sdelta}) thus yields 
$\widehat{\cal M}_i = s \widehat{\Delta}_i + (\widehat{{\cal C} \Delta})_i$. 
A generic  function on the zero--mode subspace
will then have the expansion
\beqa
\label{inte1}
g( \widehat{\Delta}, \widehat{\cal M}) & = & 
g_{0} (\widehat{\Delta} )+ 
g_{i_1} (\widehat{\Delta} ) \widehat{\cal M}_{i_1} + 
{1\over 2!}
g_{i_1 i_2}(\widehat{\Delta}) \widehat{\cal M}_{i_1}\widehat{\cal M}_{i_2} +
\ldots 
\nonumber \\
&+&
{1\over p!}
g_{i_1 i_2 \ldots i_p} (\widehat{\Delta}) 
\widehat{\cal M}_{i_1}\widehat{\cal M}_{i_2}\cdots \widehat{\cal M}_{i_p}
\ \ ,
\eeqa
the coefficients of the expansion being totally antisymmetric in their indices.
Now the first of (\ref{azzarolina}) 
implies that the
$\widehat{\cal M}_{i}$'s and the $s\widehat{\Delta}_{i}$'s are related 
by a (moduli--dependent) linear transformation $K_{ij}$, 
which is completely known 
once the explicit expression for ${\cal C}$ is 
plugged into the $\widehat{\cal M}_{i}$'s:
\beq
\label{inte2}
\widehat{\cal M}_{i} = K_{ij} ( \widehat{\Delta} ) s\widehat{\Delta}_{j}
\ \ .
\eeq
It then follows that 
\beqa
\label{noncera}
\widehat{\cal M}_{i_1}\widehat{\cal M}_{i_2}\cdots \widehat{\cal M}_{i_p} &=&
K_{i_1 j_1} K_{i_2 j_2} \cdots K_{i_p j_p} 
s\widehat{\Delta}_{j_1}s\widehat{\Delta}_{j_2}\cdots s\widehat{\Delta}_{j_p} =
\nonumber \\
&=& \epsilon_{j_1 \ldots j_p} K_{i_1 j_1} K_{i_2 j_2} \cdots K_{i_p j_p} 
\ s^p \widehat{\Delta} =
\nonumber \\
&=&
\epsilon_{i_1 \ldots i_p} ({\rm det} K) \  s^p \widehat{\Delta}
\ \ ,
\eeqa
where $s^p \widehat{\Delta}\equiv 
s\widehat{\Delta}_1 \cdots s \widehat{\Delta}_p$.
From (\ref{inte1}), (\ref{inte2}) we conclude that
\beqa
\int_{\mathscr{M}^+}g( \widehat{\Delta}, \widehat{\cal M}) &=& 
{1\over p!} \int_{\mathscr{M}^+} g_{i_1 i_2 \ldots i_p} (\widehat{\Delta} ) 
\widehat{\cal M}_{i_1}\widehat{\cal M}_{i_2}\cdots \widehat{\cal M}_{i_p} =
\nonumber \\
&=&
\int_{\mathscr{M}^+}  s^p  \widehat{\Delta} \  |{\rm det} K| 
g_{1 2 \ldots p} (\widehat{\Delta} )
\ \ .
\eeqa
This  formula is an operative tool to calculate physical amplitudes.
Here the determinant of $K$ naturally stands out as 
{\it the instanton integration measure for $N=2$ SYM theories.} 
This important ingredient of the calculation
is obtained in standard instanton calculations as
a ratio of bosonic and fermionic zero--mode Jacobians.
Instead, in our approach it emerges in a geometrical 
and  very direct way, 
{\it without the  need  of any computation of 
ratios of determinants,  nor of any knowledge of 
the explicit expressions of bosonic and fermionic zero--modes.}
The only ingredient is  the connection  ${\cal C}$.
As an instructive exercise, in the following subsection
we will compute $K$ and its determinant (\ie\ the instanton measure)
in the cases
of winding number equal to one and two. We anticipate that 
the results we  get will agree with previously known formulae; 
however they  are obtained here in a very quick and straightforward way. 

A last remark concerns what happens to the action 
of the theory,  $S_{\rm TYM}$,  when it is restricted to 
the zero--mode subspace (we called the corresponding expression
$S_{\rm inst}$ for obvious reasons).  In sec.\,\ref{venanzio}  we saw
that for $v\neq 0$ it is non--vanishing; its expression was given in  
(\ref{DKM}). In the following we 
will need to explicitly compute $S_{\rm inst}$ 
as a function of the instanton moduli; 
this will be done in sec.\,\ref{puzzola}, where  we 
will also be able to write   it as a total BRST derivative. 

\subsection{The Instanton Measure for Winding Number $k=1,2$}
\label{victor}
The ADHM bosonic and fermionic matrices can be written as 
\beq
\label{matricik=1}
\Delta=\pmatrix{w\cr x_0 -x}\ \  , \ \ 
{\cal M}=\pmatrix{\mu \cr \xi}\ \ .
\ee
For $k=1$ there are no constraints over the collective coordinates;
therefore the left--over reparametrization 
group introduced in (\ref{nonbanane}) is trivial. 
As a result we simply have  
\beq
\label{alan}
{\cal M} =\pmatrix{s w\cr s x_0 }\ \  ,
\eeq
and ${\rm det }K = 1$. The instanton measure is then given by 
$s^4x_0 s^4 w$, which is the well--known 't Hooft measure \cite{th}.
We now move on to the more interesting case of $k=2$.

The ADHM bosonic matrix reads
\beq
\Delta=\pmatrix{w_1 & w_2\cr x_1 - x & a_1\cr a_1 & x_2 - x}
=\pmatrix{w_1 & w_2\cr a_3 & a_1\cr a_1 & -a_3}+b(x-x_0) \ \ ,
\label{f.14}
\eeq
where $x_0=(x_1+x_2)/2$, $a_3=(x_1-x_2)/2$. 
We also need the expression of the matrix ${\cal M}$ 
which is defined in (\ref{fconstr}). Since this constraint 
is very similar to (\ref{bos}) (to get convinced of 
this fact just think that two solutions of (\ref{fconstr}) are given 
by ${\cal M}$ proportional to $a$ and $b$)
it is convenient to choose a form of ${\cal M}$ which parallels 
(\ref{f.14}), \ie\
\be
{\cal M}=\pmatrix{\mu_1 & \mu_2\cr \xi+{\cal M}_3 & {\cal M}_1\cr {\cal M}_1
& \xi-{\cal M}_3}
=\pmatrix{\mu_1 & \mu_2\cr {\cal M}_3& {\cal M}_1\cr 
{\cal M}_1& -{\cal M}_3}- b\xi\ \ .
\label{f.155}
\ee
The solution to the bosonic constraint (\ref{bos}) is simply given by  
\beq 
\label{natale}
a_1=\frac{1}{4|a_3|^2}a_3(\bar{w}_2 w_1-\bar{w}_1w_2 +\Sigma)\ \ ,
\eeq
where $\Sigma$ is an arbitrary real parameter related
to  the left--over $O(2)$ symmetry. 
In the following we  will exploit this $O(2)$ gauge freedom to put 
$\Sigma$   to zero.
The constraint (\ref{fconstr}) is satisfied imposing
\be
{\cal M}_1=\frac{a_3}{2|a_3|^2}(2\bar{a}_1 {\cal M}_3+
\bar w_2\mu_1-\bar w_1\mu_2)
\ \ ;
\label{f.16}
\ee
from now on
we will choose $\left\{ \mu_1, \mu_2 , \xi, {\cal M}_3 \right\}$ as a set
of independent fermionic variables.
Finally, the equation $L\cdot{\cal A}'=-\Lambda_b-\Lambda_f$ reduces to
 \beqa
H({\cal A}'_f)_{12}&=&(\Lambda_f)_{12}
\equiv\bar{\mu}_1 \mu_2-\bar{\mu}_2 \mu_1+
2(\bar{\cal M}_3{\cal M}_1-\bar{\cal M}_1{\cal M}_3)\ \ ,
\\ 
H({\cal
 A}'_b)_{12}&=&(\Lambda_b)_{12}\equiv
\bar{w}_1{\cal A}_{00}w_2-\bar{w}_2{\cal A}_{00}w_1
\ \ ,
\eeqa
where $H=|w_1|^2+|w_2|^2+4(|a_3|^2+|a_1|^2)$. 
Let us now write the BRST transformations of the bosonic ADHM
 variables:
\beqa
\label{brsk=2}
\mu_1&=&sw_1+{\cal C}_{12} w_2+{\cal C}_{00} w_1
\ \ ,
\nonumber\\
\mu_2&=&sw_2-{\cal C}_{12} w_1+{\cal C}_{00} w_2\ \ ,\nonumber\\
\xi&=&sx_0\ \ ,\nonumber\\
{\cal M}_3&=&sa_3+2\ {\cal C}_{12} a_1\ \ ,\nonumber\\ 
{\cal M}_1&=&sa_1-2\ {\cal C}_{12} a_3\ \ .
\eeqa
The component ${\cal C}_{12}$ of the $O(2)$ connection 
\beq
{\cal C}^\prime =\pmatrix{0 &  {\cal C}_{12}\cr  - {\cal C}_{12} & 0 }
\ \ 
\eeq
can be simply obtained 
by plugging the right hand sides of (\ref{brsk=2}) into the fermionic
constraint 
$(\Delta^{\dagger}{\cal M})_{12}=(\Delta^{\dagger}
{\cal M})_{21}$ and solving for ${\cal C}_{12}$.  
Actually, the terms containing ${\cal C}_{00}$, 
which is given by (\ref{C00}),  can be discarded,
since they do not contribute upon integration on the instanton 
moduli space. 
This way we get 
\beq
\label{connk=2}
{\cal C}_{12}=\frac{1}{H}\bigg[
\bar{w}_1 sw_2 -
\bar{w}_2 sw_1+ 2(\bar{a}_3 sa_1-\bar{a}_1 sa_3)\bigg]
\ \ .
\eeq
Eliminating $sa_1$ via  (\ref{natale}) 
(in the gauge $\Sigma = 0$),  one can rewrite ${\cal C}_{12}$
in terms of differentials of independent bosonic moduli, thus obtaining 
\beqa
\label{connk=2/2}
{\cal C}_{12}&=&\frac{1}{2H}\bigg[
\bar{w}_1 sw_2-
\bar{w}_2 sw_1
- 4 \bar{a}_1 s a_3 +
\nonumber \\
&+& 
s\bar{w}_2 w_1 - 
s\bar{w}_1 w_2 
- 4 s \bar{a}_3  a_1
\biggr]
\ \ .
\eeqa

Two observations are in order. 
First, we remark that 
(\ref{connk=2/2}) clearly shows that ${\cal C}_{12}$ is real,
as a connection of an orthogonal group should. 
Moreover, the r.h.s. of (\ref{connk=2/2})  does not depend on $sx_0$;
for this reason and from (\ref{brsk=2}) it immediately follows 
that in computing ${\rm det} K$ we can discard  the variable 
$\xi\equiv sx_0$,
which would contribute with the determinant of a unit matrix.
We will then define a ``reduced" fermionic matrix (of
quaternions) $\widetilde{\cal M}$ as 
\beq
\label{cocco}
\widetilde{\cal M}=\pmatrix{\mu_1\cr
\mu_2\cr
{\cal M}_3}
\ \ ,
\eeq
its bosonic  counterpart being 
\beq
\label{boscounter}
\widetilde{ \Delta }=\pmatrix{w_1\cr
w_2\cr
a_3}
\ \ .
\eeq
The relation between 
$\widetilde{\cal M}$ and $s \widetilde{\Delta}$ can be cast into the form 
\beq
\label{cappa}
({\widetilde{\cal M}}_{\alpha\dot{\alpha}})_i=
\sigma^{\mu}_{\alpha\dot{\alpha}}(K_{\mu\nu})_{ij}(s\widetilde{\Delta}_{\nu})_j
\ \ ,
\eeq
where $i=1,2,3$.
Plugging (\ref{connk=2/2}) into (\ref{brsk=2}), 
we get, after a little algebra, 
the following explicit expression for $K$, 
\beq
\label{trbosfer}
(K_{\mu\nu})_{ij}=\pmatrix{\delta_{\mu\nu}-w_{2\mu}w_{2\nu}/ H&
w_{2\mu}w_{1\nu}/ H&
-4 w_{2\mu}a_{1\nu}/ H\cr\cr
w_{1\mu}w_{2\nu}/ H&\delta_{\mu\nu}-w_{1\mu}w_{1\nu}/ H&
4 w_{1\mu}a_{1\nu}/ H\cr\cr
-2 a_{1\mu}w_{2\nu}/ H&2 a_{1\mu}w_{1\nu}/ H&
\delta_{\mu\nu}-8a_{1\mu}a_{1\nu}/ H}
\ \ ,
\eeq
whose determinant we want now to compute.

To this end, let us write $K$  as
\beq
K=\uno-z z^T P=(P^{-1}-z z^T)P\ \ ,
\eeq
where 
\beq
P=\pmatrix{\uno&0&0\cr0&\uno&0\cr0&0&2\cdot\uno}
\ \ ,
\eeq
and 
\beq
z=\frac{1}{\sqrt{H}}\pmatrix{w_2\cr -w_1\cr 2a_1}\ \ .
\eeq 
It is easy to verify that the determinant of a matrix 
of the form 
\beq
Q={\rm diag}(\alpha_1,\ldots,\alpha_n)- z z^T \ \ , \ \  z=
\pmatrix{z_1\cr\vdots \cr z_n}\ \ ,
\eeq
is simply 
\beq
\det Q=\prod_{i=1}^{n}\alpha_i-\sum_{i=1}^{n}\left(\prod_{j\neq i}
\alpha_j\right)|z_i|^2\ \ ,
\eeq
from which it is straightforward to get 
\beq
\label{jaco}
|\det K|=\frac{4\Big| |a_3|^2-|a_1|^2 \Big|}{H} 
\ \ . 
\eeq
Restoring the $\Sigma$ dependence of $a_1$ and $sa_1$ in
(\ref{connk=2/2}) we obtain 
\beqa
{\cal C}_{12}&=&\frac{1}{2H}\bigg[
\bar{w}_1 sw_2-
\bar{w}_2 sw_1
- 4 \bar{a}_1 s a_3 +
\nonumber \\
&+& 
s\bar{w}_2 w_1 - 
s\bar{w}_1 w_2 
- 4 s \bar{a}_3  a_1+ s\Sigma
\biggr]
\ \ ,
\eeqa
where $\Sigma=\Sigma(w_1,w_2,a_3,x_0)$. ${\cal C}_{12}$ contains
also a term proportional to $sx_0$; however, this term turns out not
to contribute to $\det K$ and, in fact, we find
\beq\label{detsigma}
|\det K|=\frac{4}{H}\left|
|a_3|^2-|a_1|^2+\frac{1}{4}\frac{\p\Sigma}{\p a_{3\m}}a_{1\m}+
\frac{1}{8}\frac{\p\Sigma}{\p w_{1\m}}w_{2\m}-
\frac{1}{8}\frac{\p\Sigma}{\p w_{2\m}}w_{1\m}\right|\ \ .
\eeq
It is now possible to write the terms containing $\Sigma$ in a more
compact way. To this end, recall that
the action of the $O(2)$ reparametrization group on 
$(w_1,w_2,a_3,a_1)$ can be read from 
(\ref{nonbanane}) and (\ref{ahoo}) with $q=\uno$ and 
\beq
R=\pmatrix{\cos\theta&\sin\theta\cr -\sin\theta&\cos\theta}\ \ .
\eeq
It is straightforward to show that 
\beqa
a_{1\m}&=&\left.-\frac{1}{2}\frac{\p a_{3\m}^{\theta}}
{\p\theta}\right|_{\theta=0}\ \ ,\nonumber\\
w_{2\m}&=&\left.-\frac{\p w_{1\m}^{\theta}}
{\p\theta}\right|_{\theta=0}\ \ ,\nonumber\\
w_{1\m}&=&\left.\frac{\p w_{2\m}^{\theta}}
{\p\theta}\right|_{\theta=0}\ \ ,
\eeqa
so that we can finally rewrite (\ref{detsigma}) as
\beq
|\det K|=\frac{4\left|
|a_3|^2-|a_1|^2-\left.\frac{1}{8}\frac{\p\Sigma^{\theta}}{\p\theta}
\right|_{_{\theta=0}}\right|}{H}\ \ .
\eeq
This result exactly gives  the instanton measure for 
$N=2$ pure SYM, which was  obtained in \cite {DKM} as 
a ratio of fermionic and bosonic Jacobians.
In our formalism it is simply the determinant of a 
coordinate transformation, and it is possible to write it down 
with the precise knowledge of  ${\cal C}$ alone. 
We will further clarify the r\^{o}le of  ${\cal C}$ in the last 
section of this work, where we will construct the moduli
space of self--dual gauge connections as a hyperk\"ahler quotient
space.

\subsection{The Multi--Instanton Action from the TYM Action}
\label{puzzola}
When restricted to the zero--mode subspace, the TYM action vanishes up
to the boundary term written in (\ref{DKM}), 
which leads to the  non--trivial multi--instanton action 
\beq
\label{azione}
S_{\rm inst}
=s\int\ d^4 x \ 2\, \partial^{\mu}\Tr(\bar{\phi}\psi_{\mu})
=4\pi^2 \lim_{|x|\to\infty} |x|^3 \frac{x^{\mu}}{|x|}s\Tr(\bar{\phi}
\psi_{\mu})\ \ .
\eeq
In the limit $|x|\to\infty$, the only non--vanishing term is given
by $\Tr[\bar{\phi}(s\psi_{\mu})]$. Let us now calculate the asymptotic
limits of $\bar{\phi}$ and $\psi_{\mu}$ for winding number $k$.
For the scalar field we trivially have 
$\lim_{|x|\to\infty}\bar{\phi}=\bar{\cal A}_{00}$,  while for
$\psi_{\mu}$ 
\beq
\lim_{|x|\to\infty}\psi_{\mu}=\lim_{|x|\to\infty}
\bigg(U^{\dagger}{\cal M}fb\bar{\sigma}_{\mu}U-
U^{\dagger}b\sigma_{\mu}f{\cal M}^{\dagger}U\bigg)\ \ .
\eeq
Knowing that asymptotically 
\beqa 
U_{0}&\to&\sigma_0\ \ ,\nonumber\\
U_{k}&\to&-\frac{1}{|x|^2}x\bar{w}_{k}U_{0}\ \ , \\
f_{kl}&\to&\frac{1}{|x|^2}\delta_{kl}\ \ ,\nonumber
\eeqa
we get 
\beqa
\label{spsiasin}
\lim_{|x|\to\infty}x^{\mu}\psi_{\mu}&=&
\lim_{|x|\to\infty}x^{\mu}
\sum_{p,l,m=1}^{k}\bigg[U^{\dagger}_0 \mu_{p}
\left(\frac{1}{|x|^2}\delta_{pl}\right)b_{lm}\bar{\sigma}_{\mu}U_{m}+
\nonumber\\
&&-
U^{\dagger}_p b_{pl}\sigma_{\mu}
\left(\frac{1}{|x|^2}\delta_{lm}\right)\bar{\mu}_{m}U_{0}
\bigg]=
\nonumber\\
&=&\frac{1}{|x|^2}\sum_{l=1}^{k}(\mu_{l}\bar{w}_{l}-
w_{l}\bar{\mu}_{l})
\ \ ;
\eeqa
from this we conclude that 
\beqa
\label{azione2}
[S_{\rm inst}]_k &=& 
4\pi^2 s \Tr \left[ 
\bar{\cal A}_{00} \sum_{l=1}^{k}
\left( \mu_{l}\bar{w}_{l}- w_{l}\bar{\mu}_{l}\right)\right]
\nonumber \\
&=&4\pi^2 s\Tr \left[\bar{\cal A}_{00}({\cal M}\Delta^{\dagger}-
\Delta{\cal M}^{\dagger})_{00}\right]=
-4\pi^2 s \Tr \left[ 
\bar{\cal A}_{00} 
\left( \Delta \ccc
\Delta^\dagger
\right)_{00}\right]
\ \ ,
\eeqa
which explicitly gives the instanton action as the total BRST
variation of a function of the bosonic and fermionic collective coordinates.
In the second equality, the subscript $00$ stands for the 
upper left entry of the matrix in parentheses, and
${\cal S}$ is the covariant derivative on instanton moduli space
defined in (\ref{dieci}).
Note that only some moduli are involved in this expression, more
precisely, the unconstrained ones.

It is easy to convince oneself that (\ref{azione2}) 
reproduces the instanton action for
$N=2$ SYM as written in \cite{DKM}. To this end,  let us now 
act with the operator $s$  on the moduli. 
From (\ref{azzarolina}) and  (\ref{tothefacc}) it follows that 
\beqa
s w_l &=& 
{\mu}_{l} - ( {\cal C}_{00}w_l -  \sum_{p=1}^{k}w_p {\cal C}^\prime_{pl})
\ \ ,
\\ \nonumber
s{\mu}_{l} &=&
{\cal A}_{00}w_l -  
\sum_{p=1}^{k}w_p {\cal A}^\prime_{pl} - {\cal C}_{00}{\mu}_{l}
\ \ ;
\eeqa
we get  
\beqa
\label{yassen}
s \Tr \left[ 
\bar{\cal A}_{00} \sum_{l=1}^{k}
\left( \mu_{l}\bar{w}_{l}- w_{l}\bar{\mu}_{l}\right)\right]
& = & 
\Tr \bigg[
2(\bar{\cal A}_{00} {\cal A}_{00}) \sum_{l=1}^{k} |w_l|^2-
2\bar{\cal A}_{00} \sum_{l=1}^{k} \mu_{l}\bar{\mu}_{l} +
\nonumber \\
&&
- 2\bar{\cal A}_{00} \sum_{l,p=1}^{k} w_l{\cal A}^{\prime}_{lp}\bar{w}_p 
+
\nonumber\\
&&
+ \sum_{l=1}^{k}
\left( 
-\bar{\cal A}_{00}{\cal C}_{00} + {\cal C}_{00} \bar{\cal A}_{00} \right)
\mu_{l} \bar{w}_l +
\nonumber\\
&&
+ \sum_{l=1}^{k}
\left( \bar{\cal A}_{00}{\cal C}_{00} -{\cal C}_{00} \bar{\cal A}_{00} \right)
w_l \bar{\mu}_{l}  
\ \ ,
\eeqa
where we also used the fact that $s{\cal A}_{00} = 0$.
The last  two terms in (\ref{yassen}) vanish by virtue of 
(\ref{C00}); therefore,  we conclude that
\beq
\label{azione3}
\left[ S_{\rm inst}\right]_{k}=
4\pi^2\Tr\bigg[
2(\bar{\cal A}_{00}{\cal A}_{00}) \sum_{l=1}^{k}|w_{l}|^2-2\bar{\cal A}_{00}
\sum_{l=1}^{k}\mu_{l}\bar{\mu}_{l}+
\sum_{l,p=1}^{k}(\bar{w}_{l}\bar{\cal A}_{00}
w_{p}-\bar{w}_{p}\bar{\cal A}_{00}w_{l}){\cal A}^{\prime}_{lp}\bigg]
\eeq
which exactly reproduces the $N=2$ SYM action in
moduli space%
\footnote{Note that in our notations 
$2\Tr ( \bar{\cal A}_{00}{\cal A}_{00}) = |v|^2$.}  \cite{DKM}. 
$\left[ S_{\rm inst}\right]_{k}$ can be decomposed as 
$\left[ S_{\rm inst}\right]_{k}= 
\left[S_B\right]_{k} + \left[S_F\right]_{k}$, where
$\left[S_{B}\right]_{k}$ ($\left[S_{F}\right]_{k}$) is the 
Higgs action  (the Yukawa action) 
for instanton number $k$. Explicitly
\beqa
\label{abm}
\left[ S_{B}\right]_{k}&=&
4\pi^2\Tr\bigg[
2(\bar{\cal A}_{00}{\cal A}_{00}) \sum_{l=1}^{k}|w_{l}|^2+
\sum_{l,p=1}^{k}(\bar{w}_{l}\bar{\cal A}_{00}
w_{p}-\bar{w}_{p}\bar{\cal A}_{00}w_{l})({\cal A}^{\prime}_b)_{lp}\bigg]
\ \ ,
\\
\label{afm}
\left[ S_{F}\right]_{k}&=&
4\pi^2\Tr\bigg[ -2\bar{\cal A}_{00}
\sum_{l=1}^{k}\mu_{l}\bar{\mu}_{l}+
\sum_{l,p=1}^{k}(\bar{w}_{l}\bar{\cal A}_{00}
w_{p}-\bar{w}_{p}\bar{\cal A}_{00}w_{l})({\cal A}_{f}^{\prime})_{lp}\bigg]
\ \ ,
\eeqa
${\cal A}_{b}^{\prime}$ and ${\cal A}_{f}^{\prime}$ being defined in 
(\ref{decca}).

We are now ready to perform explicit instanton calculations in our 
framework. 
\section{Computation of Instanton--Dominated 
Correlators in the Seiberg--Witten Model}
\setcounter{equation}{0}
\label{marcella}
The strategy for computing instanton--dominated correlators in our
set--up has been described at the beginning of sec.\,\ref{ilaria}.
Here we focus the attention on the Green's function 
$\left< \Tr \phi^2 \right>$, 
which is relevant for the computation of the Seiberg--Witten
low--energy effective action\cite{mat,ft}. 
To begin with, notice that the group of translations in $\Rq$
is a symmetry of the theory even in the case $v\neq 0$; as a consequence 
$x_0$ and its  supersymmetric counterpart
$s x_0 \equiv \xi$ (which is naturally expressed 
as the BRST variation of the instanton configuration center $x_0$) 
do  not appear in 
$S_{\rm inst}$, as a direct check of (\ref{azione3}) also shows. 
They will then have to be soaked up by selecting the translational
part in the correlator
insertion $\Tr \phi^2$; this amounts to performing the replacement
$\phi\longrightarrow F_{\mu \nu} \cdot (1/2)  (s x_{0\mu}s x_{0\nu})$,
\ie
\be
\Tr \phi^2 
\longrightarrow
{1\over 2}  
\Tr (F_{\mu \nu}\tilde{F}_{\mu \nu})\ s^4x_0
\ \ .
\label{fireplace}
\ee
The integral over these collective coordinates can now be easily performed 
giving the winding number \cite{ft},
\be 
\int_{ \{ x_0 \} } {\Tr \phi^2 \over 8\pi^2}  \longrightarrow 
- k \ \ .
\eeq
Therefore,  we get 
\beq
\label{elena}
<{\Tr\phi^2\over 8\pi^2}>_{k}
=
- k \int_{\mathscr{M}^+\backslash \{x_0\}}
e^{- [S_{\rm inst}]_{k}}
\ \ .
\eeq

The last step consists in integrating the exponential of the instanton  
action over the remaining collective coordinates, \ie\  
over    the  ``reduced''  moduli space 
$\mathscr{M}^+\backslash \{x_0\}$, whose   dimension  
is  $4n$ where $n=2k-1$. 
Let us call $\widetilde{\Delta}_i$, 
$i=1,\ldots,n$   the ADHM data
for $\mathscr{M}^+\backslash \{x_0\}$,   
and $\widetilde{\cal M}_i$, $i=1,\ldots,n$
their fermionic counterpart (therefore $\xi$ is not  included in  the 
$\widetilde{\cal M}_i$'s). The $\widetilde{\Delta}_i$'s and
the $\widetilde{\cal M}_i$'s 
are respectively the generalizations 
of (\ref{boscounter})  and (\ref{cocco}) 
for instanton number $k$.
After substituting the solutions to the 
fermionic constraint  (\ref{fconstr})
in (\ref{afm}), 
$[S_F]_k$ can be written as 
\beq
\label{yukact}
[S_F]_k=
\overline{\widetilde{\cal M}}_{i}^{\dot{A}\alpha}
(h_{ij})_{\alpha}{}^{\beta}
(\widetilde{\cal M}_{j})_{\beta\dot{A}}
\ \  ,
\eeq
where $i,j=1,\ldots,n$ and%
\footnote{In the following equation  we denote by $h^\dagger$ 
the hermitean conjugate matrix  obtained {\it without} complex conjugating
$v$, {\it i.e.} treating $v$ as real.}
$h=-h^\dagger$; 
let us also define 
$h_{ij}= 8\pi^2 \widehat{h}_{ij}$, for the sake of future
convenience.
In the $k=1$ case one simply has $\widehat{h}= \bar{v}$, whereas for
$k=2$ the explicit expression for  $\widehat{h}_{ij}$ will be written
in (\ref{germani}).

The exponential of $[S_F]_k$ can now be expanded in powers. 
Under the integration over the reduced moduli space 
$\mathscr{M}^+\backslash \{x_0\}$ the only surviving 
term of the expansion will be  the one that, after using (\ref{eccheneso}),  
produces the top form on $\mathscr{M}^+\backslash \{x_0\}$.
It is crucial to remark that all the terms containing  ${\cal C}_{00}$
do not contribute to the amplitudes
since the parameter $\lambda$ introduced in  (\ref{brstZ-op}) does not 
belong to the moduli space. In order to better perform this expansion, 
let us define 
\beq
\widetilde{\cal M}_i=
\pmatrix{
(\widetilde{\cal M}_i)_4+i(\widetilde{\cal M}_i)_3
&
i(\widetilde{\cal M}_i)_1+(\widetilde{\cal M}_i)_2
\cr
i(\widetilde{\cal M}_i)_1-(\widetilde{\cal M}_i)_2
&
(\widetilde{\cal M}_i)_4-i(\widetilde{\cal M}_i)_3
\cr}
=
\pmatrix{(\eta_i)_1&-(\bar\eta_i)_2\cr(\eta_i)_2&(\bar\eta_i)_1\cr}
\ \ , 
\eeq
where
$(\widetilde{\cal M}_i)_\mu$, $\mu=1,\ldots,4$ are 
the Cartesian components of $\widetilde{\cal M}_i$.
(\ref{yukact}) can then be written as 
\beqa
\label{yukact1}
[S_F]_k&=&
\bar{\eta}_{i}^{\alpha}
\left[
(h_{ij})_{\alpha}{}^{\beta}
+ \epsilon_{\alpha \delta}
(h_{ji})_{\sigma}{}^{\delta}
\epsilon^{\beta \sigma}
\right]
(\eta_{j})_{\beta}= 
\bar{\eta}_{i}^{\alpha}
[(h- h^\dagger)_{ij}]_{\alpha}{}^{\beta}
(\eta_{j})_{\beta}
\nonumber \\
&=& 
2\bar{\eta}_{i}^{\alpha}
(h_{ij})_{\alpha}{}^{\beta}
(\eta_{j})_{\beta}
\ \ ,
\eeqa 
since $h$ is anti--hermitean. 
In order to recognize the coefficient of the top form, 
we now
explicitly expand $\exp{(-[S_F]_k)}$.   
After a little algebra,  one finds 
\beq
\label{grassint}
\left. e^{-[S_F]_k}\right|_{top\ form}=
(32\pi^2)^{2n}\det \widehat{h} 
\prod_{i=1}^{n}
\left[ 
(\widetilde{\cal M}_i)_1
(\widetilde{\cal M}_i)_2
(\widetilde{\cal M}_i)_3
(\widetilde{\cal M}_i)_4
\right]
\ \ ,
\eeq
where we used (\ref{yukact1}) and 
$\eta_1\eta_2\bar{\eta}_{1}\bar{\eta}_{2}=
-4\widetilde{\cal M}_1\widetilde{\cal M}_2
\widetilde{\cal M}_3\widetilde{\cal M}_4$.
The coefficient of the top form on the 
reduced  moduli space is then proportional to the determinant 
of the matrix $H$. However,  one more ingredient now emerges:
the matrix $K$ of the change of coordinate basis between 
${\cal M}_i$ and $s\Delta_i$;
recalling (\ref{noncera}), we  conclude in fact that
\beq
\label{grassint2}
\left. e^{-[S_F]_k}\right|_{top\ form}=
(32\pi^2)^{2n}\det \widehat{h} |\det K | \ s^{4n}\widetilde{\Delta}
\ \ .
\eeq
Finally, inserting (\ref{grassint2}) in (\ref{elena}) we get 
\beq
\label{tamara3}
<{\Tr\phi^2\over 8\pi^2}>_{k}
=
-k \cdot (32\pi^2)^{2n}  
\int_{\mathscr{M}^+\backslash \{x_0\}}s^{4n}\widetilde{\Delta}\ 
\det \widehat{h} |\det K | e^{-\left[S_{B}\right]_{k}}
\ \ .
\eeq
where $[S_B]_k$ is written  in (\ref{abm}).
Note that in (\ref{grassint2}), (\ref{tamara3}) 
the instanton integration measure 
$|\det K|$ has naturally come out.

This is our starting point. Let us now perform 
the $k=1$ computation explicitly.

\subsection{The $k=1$ Case in the Bulk}
\label{bob!}
In the $k=1$  case, 
the structure of the instanton moduli  space $\mathscr{M}^+$
has been thoroughly investigated, and 
it is explicitly known to be the manifold
$\R^4\times \R^{+}\times S^3/\Z_2$
\cite{dk,Freed}; the three factors correspond respectively
to the instanton center ($x_0$), scale ($|w|$) and 
orientation in color space ($w/|w|$). 
The reduced moduli space $\mathscr{M}^+\backslash \{x_0\}$ 
is then the 4--dimensional manifold 
$\R^{+}\times S^3/\Z_2$,  obtained
after first integrating out the collective coordinate $x_0$ \cite{ft}. 

The ADHM bosonic and fermionic matrices are written in (\ref{matricik=1}), 
and the action (\ref{azione3}) 
calculated on the one--instanton background is given
by
\beq
\label{ack=1}
\left[S_{\rm inst}\right]_{k=1} = 
4 \pi^2 \left[ |v|^2|w|^2  
- 2\Tr (  \bar{v}\mu \bar{\mu} ) \right]
\ \ . 
\eeq 
Taking into account that
\beq
\Tr(\mu\bar\mu\bar v) =
-sw_\mu sw_\nu \sum_{a=1}^{3}\eta^a_{\mu\nu}(v^{a})^*
\ \ ,
\eeq
where $\eta^a_{\mu\nu}$ are the 't Hooft symbols, 
we get, after a little algebra,
\beqa
\label{tamara}
<{\Tr\phi^2\over 8\pi^2}>_{k=1}
&=&
- \int_{\mathscr{M}^+\backslash \{x_0\}}
e^{- [S_{\rm inst}]_{k=1}}=
- \int_{\mathscr{M}^+\backslash \{x_0\}}
e^{-4 \pi^2 \left[ |v|^2|w|^2  
- 2{\rm Tr} (\mu \bar{\mu}\bar{v} )\right]}
\nonumber\\
&=&
- {(8\pi^2)^2\over 2!}
\int_{\mathscr{M}^+\backslash \{x_0\}}
e^{-4\pi^2 |v|^2|w|^2}\Tr(\mu\bar\mu\bar{v})
\Tr(\mu\bar\mu\bar{v})\nonumber\\
&=&
- (8\pi^2)^2\cdot 2^2  \int_{\mathscr{M}^+\backslash \{x_0\}} (v^{*})^2
e^{-4\pi^2 |v|^2|w|^2}s^4w = - {8\pi^2\over v^2}
\ \ ,
\eeqa
which is the expected result \cite{ft}.
In (\ref{ack=1}) and hereafter 
we use the shorthand notation $\bar{v}=\bar{\cal A}_{00}= -v^*\sigma^3 /2i$. 

\subsection{The $k=2$ Case in the Bulk}
\label{7777}
In this subsection we will describe the $k=2$ computation in the bulk
{\it i.e.} without using the property that the action is BRST exact. 
All the features of the topological approach will now become apparent.

The action (\ref{azione3})
calculated on the two--instanton background is given
by
\beqa
\label{ack=2}
\left[S_{\rm inst}\right]_{k=2}
&=&
[S_B+S_F]_{k=2}=4\pi^2|v|^2(|w_1|^2+|w_2|^2)-16\pi^2{|\omega|^2 
\over H}
\\
&+&8\pi^2\Tr\left\{\bar{\mu}_1 \bar{v}\mu_1+\bar{\mu}_2 
\bar{v}\mu_2+
\frac{\bar{\omega}}{H}\left[\bar{\mu}_1 \mu_2-\bar{\mu}_2 \mu_1+
2(\bar{\cal M}_3{\cal M}_1-\bar{\cal M}_1{\cal M}_3)\right]\right\}
\ \ , 
\nonumber
\eeqa 
where we have defined 
\be
\omega=-\Lambda_{b12}({\cal A}_{00}) = 
- \bar{w}_1{\cal A}_{00}w_2+\bar{w}_2{\cal A}_{00}w_1
\ \ .
\eeq
After substituting the fermionic constraint (\ref{f.16}) in (\ref{ack=2}),
$[S_F]_{k=2}$ can be written as in 
(\ref{yukact}); the indices 
$i,j$ run from 1 to 3, 
the $\widetilde{\cal M}_{i}$'s are defined in (\ref{cocco}), and   
$h_{ij}= 8\pi^2 \widehat{h}_{ij}$, where explicitly     
\beq
\label{germani}
\widehat{h}_{ij}=\pmatrix{\bar{v}&\frac{\bar{\omega}}{H}&
-\frac{\bar{\omega}}{H|a_3|^2}w_2\bar{a}_3\cr\cr
-\frac{\bar{\omega}}{H}&\bar{v}&\frac{\bar{\omega}}{H|a_3|^2}w_1
\bar{a}_3\cr\cr
\frac{\bar{\omega}}{H|a_3|^2}a_3\bar{w}_2&
-\frac{\bar{\omega}}{H|a_3|^2}a_3\bar{w}_1&
\frac{2\bar{\omega}}{H|a_3|^2}(a_3\bar{a}_1-a_1\bar{a}_3)}\ \ .
\eeq
Moreover, specializing (\ref{grassint2}) to the $k=2$ case we get 
\beq
\label{grassint22}
\left. e^{-[S_F]_{k=2}}\right|_{top\ form}=
(32\pi^2)^{6}\det \widehat{h} |\det K | s^4w_1s^4w_2s^4a_3
\ \ .
\eeq
The determinant of the matrix $K$ in 
(\ref{trbosfer}) 
was explicitly computed in (\ref{jaco}).
We want now to calculate the determinant of $\widehat{h}$.
To this end, note that 
this matrix  has the form%
\footnote{In the following equation  we denote by a bar over a quaternion
the hermitean conjugate quaternion obtained without complex conjugating
$v$; in other words, if $q\in \bb{H}$ and $v\in \C$, then   
we define $\overline{ vq} = v \bar{q}$.}
\beq
\widehat{h}=
\pmatrix{F&B&C\cr-B& F&D\cr-\bar{C}&-\bar{D}&E}\ \ ,
\eeq
where $\bar{F}=-F$ and $B=\bar{B}$.
By means of elementary operations on the rows and columns of the matrix 
$\widehat{h}$ 
({\it i.e.} by multiplying rows by quaternions and then adding and 
subtracting rows) we can write
\beq
\label{zia}
\widehat{h}=h_1h_2=\pmatrix{{F\over |F|^2|B|^2}
&{F\over |B|^2}({1\over |F|^2}-{\alpha\beta^{-1}\over |C|^2})
&-{FC\over |C|^2}\cr
0&{B\alpha\beta^{-1}\over |B|^2|C|^2}&{BC\over |C|^2}\cr0&0&1}
\pmatrix{0&0&\gamma-\beta\alpha^{-1}\delta\cr
0&\beta&\beta\alpha^{-1}\delta\cr
-\bar C&-\bar D&E}\ \ ,
\eeq
where
\beqa
\alpha&=&|C|^2\bar BF+|B|^2C\bar D\ \ ,
\nonumber\\
\beta&=&|B|^2\bar FB+|F|^2\bar BF\ \ ,
\nonumber\\
\gamma&=&|B|^2\bar FC+|F|^2\bar BD\ \ ,
\nonumber\\
\delta&=&|C|^2\bar BD-|B|^2CE\ \ .
\eeqa 
Using (\ref{zia})  one finds,  after some algebra, 
\beq
\label{detquat}
\det \widehat{h}= 
\left(\frac{\bar{\omega}v^*}{2H}\right)^2 \frac{1}{|a_3|^4}
\det\left\{\frac{v^*}{2}(\bar{w}_1 w_2-\bar{w}_2 w_1)-
\frac{i\bar{\omega}}{H}(\bar{w}_1\sigma^{3}w_1+\bar{w}_2
\sigma^{3}w_2)\right\}\ \ ,
\eeq
where $\sigma^3$ is the third Pauli matrix. 
(\ref{detquat}) is the determinant of a quaternion, {\it i.e.} the 
squared absolute value of the quaternion itself. The final result is
\beq
\label{determinant}
\det \widehat{h}= 
\left(\frac{\bar{\omega}v^*}{2H}\right)^2 \frac{1}{|a_3|^4}
\left\{2\bar{\omega}^2 \tau_1+\left(\frac{v^*}{2}\right)^2 
|\Omega|^2+\bar{\omega}^2\left[\tau_2+\frac{1}{(v^*/2)^2}\left(
\frac{\bar{\omega}}{H}\right)^2\right]\right\}\ \ ,
\eeq
where 
\beqa 
\Omega&=&w_1 \bar{w}_2-w_2 \bar{w}_1\ \ ,\nonumber\\
L&=&|w_1|^2+|w_2|^2\ \ ,\nonumber\\
\tau_1&=&\frac{L}{H}\ \ ,\nonumber\\
\tau_2&=&\frac{L^2-|\Omega|^2}{H^2}\ \ . 
\eeqa
(\ref{determinant})  reproduces 
the result known in literature \cite{DKM}.

With the aid of (\ref{grassint22}) we can 
now compute 
\beq
\label{trfi2k=2}
<{\Tr\phi^2\over 8\pi^2}>_{k=2}=
-2\cdot(32\pi^2)^{6}
\int_{\mathscr{M}^+\backslash \{x_0\}}
s^{4}w_1 s^{4}w_2 s^4{a_3} \ \det \widehat{h} |\det K|e^{-[S_B]_{k=2}}
\ \ .
\eeq
Using (\ref{jaco}) and (\ref{determinant}), 
it is  easy to see that 
the integral over the bosonic moduli which appears in 
(\ref{trfi2k=2}) 
is the same which was found in \cite{ft} after integrating out the fermionic 
zero--modes. As in \cite{ft},  
(\ref{trfi2k=2}) thus leads to 
$\left<\Tr\phi^2/  (8\pi^2)\right>_{k=2} = 
-5\cdot (8\pi^2)^3/(4v^6)$ \cite{ft}, 
which agrees with the results
found by Seiberg and Witten in \cite{sw}.
\subsection{On the Use of the Operator $s$}
\label{cracchis}
As we have observed in the previous sections, 
the operator $s$ is nilpotent.  
Moreover, it is possible to write the action $S_{\rm inst}$  
as the operator $s$ acting on a certain function of the moduli
as in (\ref{azione2}).
This enables one to write the correlator $<\Tr\phi^2>_{k}$
as an integral over the boundary of the  instanton moduli space. 
Since 
\beq
\label{act4exp}
\left[S_{\rm inst}\right]_{k}
=\left[ S_B+S_F\right]_{k}=4\pi^2 s\bigg\{\Tr\Big[\bar v(\sum_{i=1}^{k}\mu_i 
\bar w_i-w_i\bar{\mu}_i)\Big]\bigg\}\ \ ,
\eeq
and $s [S_{\rm inst}]_k=0$, we obtain
\beqa
\label{act4exp1}
e^{-[S_{\rm inst}]_k}
&=&
\sum_{l=0}^\infty {(-)^l \over l!}([S_B]_k+[S_F]_k)([S_B]_k+[S_F]_k)^{l-1}
\nonumber \\
&=&
4\pi^2 s\left\{\Tr\Big[\bar v(\sum_{i=1}^{k}\mu_i \bar w_i-w_i\bar{\mu}_i )
\Big]\sum_{l=0}^\infty {(-)^l \over l!}
([S_B]_k+[S_F]_k)^{l-1}\right\}
\\
&=&
4\pi^2 s\left\{ \Tr\Big[\bar v(\sum_{i=1}^{k}
\mu_i \bar w_i-w_i\bar{\mu}_i)\Big]
\sum_{l=0}^\infty {(-)^l \over l!}\sum_{p=0}^{l-1}
\pmatrix{ l-1 \cr p }
([S_B]_k)^{l-1-p}([S_F]_k)^p
\right\}\ \ .
\nonumber 
\eeqa
Since $[S_F]_k$ is a fermion bilinear, in order to build a fermionic top
form on the $(8k-4)$--dimensional 
reduced moduli space we must have $p=4k-3$, leading to 
\beqa
\label{act4exp2}
\left. e^{-[S_{\rm inst}]_k}\right|_{top\ form}
&=&
4\pi^2 s\left\{\Tr\Big[\bar v(\sum_{i=1}^{k}\mu_i \bar w_i-w_i\bar{\mu}_i)\Big]
{([S_F]_k)^{4k-3}\over (4k-3)!}
\sum_{l=0}^\infty {(-)^l \over (l+4k-2)}
{([S_B]_k)^{l}\over l!}\right\}
\nonumber \\
&=&4\pi^2 s\left\{\Tr\Big[\bar v(\sum_{i=1}^{k}\mu_i \bar w_i-
w_i\bar{\mu}_i)\Big] ([S_F]_k)^{4k-3}([S_B]_k)^{-4k+2}\right.
\nonumber \\
&\cdot& 
\left.\bigg(1-e^{-[S_B]_k}\sum_{l=0}^{4k-3}{([S_B]_k)^l\over l!}\bigg)\right\}
\ \  .
\eeqa
As we stated in the introduction, writing the correlator as a total derivative
over the moduli space can lead to interesting results. The $8k$--dimensional 
moduli space $\mathscr{M}_k$,
can be compactified according to \cite{dk}. 
If we denote this compactification by
$\overline\mathscr{M}_k$, 
it is well known that the boundary $\partial\overline\mathscr{M}_k$ can be 
decomposed into a union of lower moduli spaces, so that
we can write 
\beq
\overline\mathscr{M}_k=\mathscr{M}_k\cup\Rq\times\mathscr{M}_{k-1}
\cup S^2\Rq\times\mathscr{M}_{k-2}
\ldots\cup S^k\Rq
\eeq
where $S^i\Rq$ denotes the $i^{th}$ symmetric product of points of $\Rq$.
The curvature density in $S^l\Rq\times\mathscr{M}_{k-l}$ is 
\beq 
|F_k|^2=|F_{k-l}|^2+\sum_{i=1}^l 8\pi^2\delta (x-y_i)
\label{occam3}
\eeq
where $y_i\in S^i\Rq$ are the centers of the instanton.
We will check (\ref{occam3}) in the $k=2$ case using (\ref{occam1}).
Given
\beq\label{maddeche}
\Delta^{\dagger}\Delta=\pmatrix{|w_1|^2+(x_1-x)^2+|a_1|^2 &
\bar{w}_1 w_2+(\bar{x}_1-\bar{x})a_1+\bar{a}_1 (x_2-x)\cr\cr
\bar{w}_2 w_1+(\bar{x}_2-\bar{x})a_1+\bar{a}_1 (x_1-x)&
|w_2|^2+(x_2-x)^2+|a_1|^2}\ \ ,
\eeq
we observe that one part of the boundary is given by $|w_1|\to 0$. Using
(\ref{natale}) we have 
\beq
\lim_{|w_1|\to 0}\Delta\longrightarrow\pmatrix{0_{2\times 1} &\Delta_{k=1}\cr
x_1-x& 0_{1\times 1}}
\ \ ,
\eeq
and
\beq
\lim_{|w_1|\to 0}\det \Delta^{\dagger}\Delta=
(x_1-x)^2[|w_2|^2+(x_2-x)^2]
\ \ .
\eeq 
Then
\beqa\label{marimbarzi}
&&\lim_{|w_1|\to 0}\Tr (FF)_{k=2}=-{1\over 2}\lim_{|w_1|\to 0}
\Box\Box\log\det ( \Delta^{\dagger}\Delta)_{k=2} d^4x\\ \nonumber
&=&-{1\over 2}\Box\Box\log (x_1-x)^2d^4x+
\Tr(FF)_{k=1}=\Tr (FF)_{k=1}+8\pi^2 \delta^{4}(x-x_1)
\ \ .
\eeqa
Extending these  
computations to encompass all boundaries one can check (\ref{occam3}).
We leave the application of these considerations and 
of (\ref{act4exp2}) to cases 
with $k>1$ to future work and here we limit ourselves to a
simple check  of (\ref{act4exp2}) in the $k=1$ case.

From the analyses of \cite{dk,Freed}, 
it is known that the boundary of the $k=1$ moduli  space
consists of instantons of zero ``conformal'' size;
this means that if we projectively map the Euclidean flat space 
$\Rq$ onto a four sphere $S^4$, the boundary of the
corresponding transformed $k=1$ instanton moduli space 
is given by instantons of zero conformal 
size $\tau$, where $\tau$ is  
obtained from  $|w|$ through a projective 
transformation ($|w|$ itself does not represent a globally defined
coordinate on the $S^4$ instanton moduli space). 
In terms of the size $|w|$ of the $\Rq$ instanton, the 
$\tau\to 0$ limit corresponds to $|w|\rightarrow 0,\infty$. 
Specializing  (\ref{act4exp2}) to $k=1$ and inserting it in (\ref{elena}) 
we get 
\beqa
\label{k=1bordo}
<{\Tr\phi^2\over 8\pi^2}>_{k=1}
&=&
- \int_{\mathscr{M}^+\backslash \{x_0\}}
e^{-[S_{\rm inst}]_{k=1}}=
- \int_{\mathscr{M}^+\backslash \{x_0\}}
e^{-4 \pi^2 \left[ |v|^2|w|^2  
- 2{\rm Tr} (  \mu \bar{\mu}\bar{v} )\right]}
\nonumber\\
&=& - 4\pi^2 \int_{\mathscr{M}^+\backslash \{x_0\}}
s \biggl\{ 
\Tr\Big[\bar v(\mu \bar w- w\bar\mu)\Big] [S_F]_{k=1} {1\over [S_B]_{k=1}^2}
\cdot
\nonumber \\
&&\cdot
\Bigl( 1- e^{-[S_B]_{k=1}} - [S_B]_{k=1} e^{-[S_B]_{k=1}} \Bigr)\biggr\}
\nonumber \\
&=&
32\pi^4\int_{\mathscr{M}^+\backslash \{x_0\}}s\bigg\{\Tr\Big[\bar v(\mu \bar w-
w\bar\mu)\Big]
\Tr(\mu\bar\mu\bar v){1\over (4\pi^2 |v|^2|w|^2)^2}\cdot\nonumber\\
&&\cdot\Big(1-e^{-4\pi^2 |v|^2|w|^2}-4\pi^2 |v|^2|w|^2
e^{-4\pi^2 |v|^2|w|^2}\Big)\bigg\}
\ \ ,
\eeqa
where we have used (\ref{act4exp2}) with 
\beqa
\left[S_{B} \right]_{k=1} 
&=& 
4 \pi^2 |v|^2|w|^2
\ \ ,
\\ 
\left[S_{F} \right]_{k=1} 
&=& 
-8\pi^2 \Tr ( \mu \bar{\mu}\bar{v} )
\ \ . 
\eeqa 
Using Stokes' theorem, we can   compute 
$\left<\Tr\phi^2/  (8\pi^2)\right>_{k=1}$ as an integral over
the boundary  
$\partial \left( \mathscr{M}^+\backslash \{x_0\}\right) $,  which for
$k=1$ is $\partial \R^{+}\times S^3/\Z_2$. 
Since 
\beqa
\label{forms}
\Tr(\mu\bar\mu\bar v) &=&-sw_\mu sw_\nu \eta^a_{\mu\nu}(v^{a})^*=
(\sigma^a_{w} s|w|^2-|w|^2 s\sigma^a_{w})(v^{a})^*\ \ ,
\nonumber\\
\Tr[\bar v(\mu \bar w-w\bar\mu)]&=&2w_\mu 
sw_\nu\eta^a_{\mu\nu}(v^{a})^*=2|w|^2\sigma^a_{w}(v^{a})^*
\ \ ,
\eeqa
we  get 
\beq
\label{forms2}
\Tr(\mu\bar\mu\bar v) \Tr[\bar v(\mu \bar w-w\bar\mu)]=
-4 |w|^4 (v^a v^a)^* \sigma^1_w\sigma^2_w\sigma^3_w
\ \ .
\eeq
Here $\sigma^a_w$ are
the left--invariant 1--forms, defined as
$\sigma^{a}_{w}= |w|^{-2} \eta^{a}_{\mu \nu} w_\mu s w_{\nu}$,
and satisfy the relation 
$\sigma^{a}_{w} \sigma^{b}_{w} = 
\epsilon^{abc} s\sigma^{c}_{w}$. 
Plugging (\ref{forms2}) into (\ref{k=1bordo}) 
and recalling that $\int_{S^3/\bZ_2}\sigma^1_w\sigma^2_w\sigma^3_w = \pi^2$,
we get 
\beqa
<{\Tr\phi^2\over 8\pi^2}>_{k=1}&=&- \left.{8\pi^2\over v^2}
\Big(1-e^{-4\pi^2|v|^2|w|^2}-4\pi^2 |v|^2|w|^2
e^{-4\pi^2 |v|^2|w|^2}\Big)\right|_{|w|=0}^{|w|=\infty}
\nonumber \\
&=&-{8\pi^2\over v^2}
\eeqa
which is the result obtained in (\ref{tamara}).

\section{Topological Correlators in Witten's Topological Field Theory}
\setcounter{equation}{0}
\label{vittoria}
In this section we focus the attention on 
Witten's twisted formulation of $N=2$ SYM theory. 
We will put to zero the v.e.v. $v$ of the complex scalar field.
For winding number $k=1$, the top form on the
(8--dimensional) instanton moduli space is
$\Tr \phi^2 (x_1)  \Tr \phi^2  (x_2)$, and one can compute
the Green's function
$<\Tr \phi^2 (x_1)  \Tr \phi^2  (x_2)>$.
The prescription (\ref{prescription}) for computing Green's function
gives in this case
\be
\label{diego}
\left\langle\Tr \phi^2 (x_1)  \Tr \phi^2  (x_2)\right\rangle=
\int_{\mathscr{M}^+} 
\left[ 
\Tr \phi^2 (x_1) \Tr \phi^2 (x_2)
 \right]_{zero-mode\  subspace}
\ \ ,
\eeq
where we have recalled that 
the boundary term (\ref{DKM}) in $S_{\rm TYM} $ vanishes 
when $v=0$. 

We could then proceed and compute explicitly the r.h.s. of
(\ref{diego}). However, 
the observation that the BRST operator $s$ is on
$\mathscr{M}^+$  the exterior derivative 
leads us to consider, as  in subsec.\,\ref{cracchis},  
the possibility of   computing  correlators of $s$--exact operators 
as integrals of forms on the boundary of $\mathscr{M}^+$.
Indeed,  recall that we can write
\beq
\Tr \phi^2 = s K_{c}\ \ , \ \
K_c = \Tr \left(csc + {2\over 3} ccc \right) \ \ ,
\eeq
an expression which parallels the well--known relation
\beq
\Tr F^2 = d K_{A}\ \ , \ \
K_A =
\Tr \left(AdA + {2\over 3} AAA\right) \ \ .
\eeq
Using Stokes' theorem,  one is led to  re--express the r.h.s. of 
(\ref{diego}) as 
\beq
\label{gio}
\int_{\mathscr{M}^+} \Tr \phi^2 \Tr \phi^2 =
\int_{\partial \mathscr{M}^+}   K_c \Tr \phi^2 
\ \ .
\eeq
We are then faced with two different computational strategies:
\begin{itemize}
\item[\bf 1.]
the bulk calculation, and
\item[\bf 2.]
the boundary calculation.
\end{itemize}
Let us explore in detail both possibilities.
\subsection{The Calculation of $<\Tr \phi^2 (x_1)  \Tr \phi^2  (x_2)>$
in the Bulk}
From the last equation in (\ref{brstfields}) 
we know that the zero--mode configuration  for $\phi$ is 
\be
\label{aa}
\phi=U^{\dagger}{\cal M}f{\cal M}^{\dagger}U+U^{\dagger}{\cal A}U
\ \ .
\ee
The parametrization for a  
$k=1$ instanton has been described in sec.\,\ref{victor};
from this it turns out that $f (x) = (\Delta^\dagger \Delta)^{-1}=
[(x-x_0)^2 + w^2]^{-1}$.  Plugging 
the expression (\ref{alan}) for ${\cal M}$ into (\ref{aa}) 
and recalling that   ${\cal A}=0$ when  $k=1$,  we get  
\be
\phi=U^{\dagger}(s \Delta) f (s \Delta)^{\dagger}U
\ \ .
\ee
It then follows that 
\be
\label{b}
\Tr \phi^2 = 
\Tr \left[
P s \Delta f (s \Delta)^{\dagger} P s \Delta f(s \Delta)^{\dagger} 
\right]
\ \ ,
\ee
where $P$ has been introduced in (\ref{f.666}).
After a little algebra, (\ref{b}) becomes 
\be
\Tr \phi^2 (x) = -48 w^4 f^4(x) \prod_{\mu=1}^{4}
\Gamma_{\mu}(x)
\ \ ,
\ee
where the quaternionic 1--form  $\Gamma (x)$ is defined  by
\be
\Gamma (x) = sx_0 + {(x-x_0)\bar{w}\over |w|^2} sw
\ \ .
\ee
It is easy to convince oneself that
\be
\prod_{\mu=1}^{4} \Gamma_{\mu}(x_1)
\prod_{\nu=1}^{4} \Gamma_{\nu}(x_2)
= J(x_1-x_2) s^4x_0 s^4w
\ \ ,
\ee
with $J(x_1-x_2) = (x_1-x_2)^4/w^4$;
we can then write
\be
\Tr \phi^2 (x_1)  \Tr \phi^2  (x_2) = 
(48)^2  w^4 (x_1-x_2)^4 f^4(x_1)f^4(x_2) s^4x_0 s^4w
\ \ .
\ee
Plugging this expression into the r.h.s. of (\ref{diego}), 
it follows that 
\be
\label{diego2}
<\Tr \phi^2 (x_1)  \Tr \phi^2  (x_2)> = 
(48)^2(x_1-x_2)^4
\int_{\mathscr{M}^+}  s^4x_0 s^4w \ w^4  f^4(x_1)f^4(x_2) 
\ \ .
\ee
The structure of the $k=1$ moduli space
has been discussed in subsec.\,\ref{bob!}, where we have learnt that 
$\mathscr{M}^{+}_{k=1} = 
\R^4\times \R^{+}\times S^3/\Z_2$.
(\ref{diego2}) then becomes 
\be
\int_{\bR^{+}\times S^3/\bZ_2}s^4w \ w^4 
\int_{\bR^4}  s^4x_0  \ 
f^4(x_1)f^4(x_2) ={\pi^4\over 72} {1\over (x_1-x_2)^4}
\ \ ,
\label{integrand}
\ee
from which we finally get
\be
\label{maya}
<{\Tr \phi^2 (x_1) \over 8\pi^2} {\Tr \phi^2 (x_2) \over 8\pi^2} > =
{1\over 2}
\ \ .
\ee
We remark that a hasty analysis would lead to the conclusion that,
in the limit $|x_1-x_2| \rightarrow 0$, the Green's function (\ref{maya}) 
is singular due to the behavior of (\ref{integrand}). This is contrary
to the geometrical interpretation 
of this correlator as a component of the Chern class
of the bundle with curvature (\ref{effepiupsi}) \cite{bs, witten2}.
With a little more thinking 
one gets convinced that this singularity is only apparent,
as we will show in the next section. In our opinion, this interpretation of
the above--computed Green's function makes it unnatural 
the application  to it of clustering
arguments,  as recently argued  in \cite{hklm}.

We now turn to describe the same calculation performed on the  boundary 
of instanton moduli space.%
\footnote{We thank Gian Carlo Rossi for many fruitful discussions and
suggestions on
the calculations described  in the next section.}
\subsection{The Calculation of $<\Tr \phi^2 (x_1)  \Tr \phi^2  (x_2)>$
on the Boundary of $\mathscr{M}^+$}
We start off by considering (\ref{gio}), which allows us to write
\be
<\Tr \phi^2 (x_1)  \Tr \phi^2  (x_2)> = 
\int_{\partial \mathscr{M}^+}   K_c  (x_1) \Tr \phi^2 (x_2)
\ \ .
\ee
The expression  of the current  $K_c$ 
(which is a 3--form) 
is a trivial extension of   (\ref{occam2}), and reads,
for instanton number $k$,  
\be
K_c =  
\Tr \left[ P s D (s D)^\dagger D (s D)^\dagger
     + {1\over 3}  (D^\dagger s D)( D^\dagger sD)( D^\dagger sD )\right]     
\ \ .
\ee
For $k=1$ one simply has
\beq
\label{matriced}
D (x) =f^{1\over 2} (x)  \pmatrix{w\cr x_0 -x}
\ \  , 
\ee
and after a lengthy algebra, one finds
\bea
K_c (x) &=&  2f^3 (x) \biggl[
2w^4 (w^2 + 3 y^2)
\bar{\sigma}^{1}_{w}
\bar{\sigma}^{2}_{w}
\bar{\sigma}^{3}_{w}
 + 
2y^4 (y^2 + 3 w^2)
\bar{\sigma}^{1}_{y}
\bar{\sigma}^{2}_{y}
\bar{\sigma}^{3}_{y}
\\ \nonumber
&+& 2 w^2 y^2 (w^2 + y^2) 
\left( {sy^2 \over 2 y^2} - {sw^2 \over 2 w^2}\right)
(\bar{\sigma}^{a}_{y}\bar{\sigma}^{a}_{w} )
\\ \nonumber
&+&
w^2 y^2 ( y^2 - w^2)
s (\bar{\sigma}^{a}_{y} \bar{\sigma}^{a}_{w} )
\biggr]
\ \ ,
\eeqa
where we set $y=x_0-x$.
The right--invariant 1--forms $\bar{\sigma}^{a}_{z}$ are defined as
$\bar{\sigma}^{a}_{z}= |z|^{-2}\bar{\eta}^{a}_{\mu \nu} z_\mu s z_{\nu}$,
and satisfy the relation 
$\bar{\sigma}^{a}_{z}\bar{\sigma}^{b}_{z} = 
\epsilon^{abc} s\bar{\sigma}^{c}_{z}$. 
The next step consists in computing the product 
$K_c (x_1 ) \Tr \phi^2 (x_2)$. The calculation is greatly simplified if one
sets $x_1=x_2$. Moreover, 
one has to take into account only the terms that 
yield a non--vanishing result 
when integrated on the boundary of instanton moduli space.
If we do this, we get
\be
[K_c \Tr \phi^2 ](x) \longrightarrow
192 y^4 w^4 f^4 (x) 
(\bar{\sigma}^{1}_{w}
\bar{\sigma}^{2}_{w}
\bar{\sigma}^{3}_{w})
\left( {sy^2 \over 2 y^2} - {sw^2 \over 2 w^2}\right)
( \bar{\sigma}^{1}_{y}
\bar{\sigma}^{2}_{y}
\bar{\sigma}^{3}_{y} )
\ \ .
\ee
Note that $ y^4 (sy^2 / 2 y^2) 
( \bar{\sigma}^{1}_{y}
\bar{\sigma}^{2}_{y}
\bar{\sigma}^{3}_{y} ) = s^4 (x_0-x) = s^4 x_0$.
It then follows that 
\be
\int_{\partial \mathscr{M}^+}   [K_c  \Tr \phi^2] (x)= 
192\int_{S^3/\bZ_2} 
\bar{\sigma}^{1}_{w}
\bar{\sigma}^{2}_{w}
\bar{\sigma}^{3}_{w}
\lim_{|w| \to 0} \int_{\bR^4} s^4x_0\ w^4 f^4 (x)
\ \ .
\ee
Since
\bea
\lim_{|w| \to 0} \int_{\bR^4} s^4x_0\ w^4 f^4 (x)&=& 
\lim_{|w| \to 0} \int_{\bR^4} s^4x_0\ 
{w^4 \over [w^2 + (x-x_0)^2]^4 }= 
C  \int_{\bR^4} s^4x_0\ \delta^{(4)} (x-x_0) 
\cr
\cr
&=& C
\ \ ,
\eea
where
\be
C= \int_{\bR^4} s^4x\ {1\over (1+x^2)^4} = {\pi^2\over 6}
\ \ ,
\ee
we conclude that 
\be
\int_{\partial \mathscr{M}^+}   [K_c  \Tr \phi^2] (x)= 192 \pi^2 \cdot 
{\pi^2\over 6} = (8\pi^2)^2 \cdot {1\over 2}
\ \ ,
\ee
and the final result is 
\be
\label{gcr2}
< {\Tr \phi^2  \over 8\pi^2} {\Tr \phi^2  \over 8\pi^2} > =
\int_{\partial \mathscr{M}^+}   {K_c \over 8\pi^2}  
{\Tr \phi^2 \over 8\pi^2} 
= 
{1\over 2}
\ \ .
\ee
(\ref{gcr2}) coincides with the result found in (\ref{maya}). The limit 
of coincident points is thus well--defined, as we observed at the end of
the previous section.



\section{The ADHM Construction and Hyperk\"ahler Quotients}
\setcounter{equation}{0}
\label{poldo}
In this section we construct the moduli space of 
self--dual connections on flat space $\Rq$
in terms of hyperk\"ahler quotients following \cite{dk}. 
This will allow us to clarify the geometrical meaning of the
algebraic construction of the BRST transformations presented in
sec.\,\ref{algebraic}.
As explained in \cite{hklr}, the hyperk\"ahler construction of a 
quotient space can be regarded from a physicist's point of view as the gauging 
of a non--linear sigma model. The corresponding connection is
obtained in a purely geometrical way {\it directly} from the isometries of 
the constraint equation (\ref{bos}) which 
imposes the self--duality of the gauge field strength
(expressed in the ADHM formalism),  and
coincides with the connection 
${\cal C}$ introduced in sec.\ref{algebraic} and worked out explicitly in 
sec.\,\ref{victor} for the $k=2$ case. 
We will show that the square root of the 
determinant of the metric on instanton moduli space 
gives the bosonic Jacobian \cite{osbo} involved in the 
transformation of the functional integral into an integration over 
instanton moduli. 
For the construction of a gravitational instanton with this method
see \cite{bfmr},  while for an introduction to hyperk\"ahler quotients
in physicists' language  see \cite{hklr}.

The starting point is the ADHM matrix $a$,  
which for the case of $SU(2)$ 
instantons was defined in (\ref{salute}).
Actually, for the present discussion it is more 
convenient to adopt a different parametrization for $a$;
we rewrite it as
\beq
\label{salute1}
a=\pmatrix{t & s^\dagger \cr A & -B^\dagger\cr B & A^\dagger} \ \ ,
\eeq
where 
$A, B$ are  $k\times k$ complex matrices and $s, t$ are $N\times k$
and $k\times N$ dimensional matrices.  
Let us introduce the  $4k^2+4kN$--dimensional 
hyperk\"{a}hler manifold $M=\{A,B,s,t\}$. 
Given the three complex structures
$J^i_{ab}$ where $i=1,2,3$ and $a, b=1,\ldots,{\rm dim}\, M$, 
we can build the 2--forms
$\omega^i=J^i_{ab}dx^a\wedge dx^b$,  where $x$ is a choice of coordinates
on $M$. The real forms $\omega^i$ allow one to define a $(2,0)$ 
and a $(1,1)$ form
\beqa
\label{comforms}
\omega_{\bC}&=& \Tr\, dA\wedge dB+\Tr\, ds\wedge dt\ \ ,\nonumber\\
\omega_{\bR}&=&\Tr\, dA\wedge dA^\dagger+\Tr\, dB\wedge dB^\dagger+
\Tr\, ds\wedge ds^\dagger- \Tr\, dt^\dagger\wedge dt\ \ .
\eeqa
The transformations
\beqa\label{invmod}
A&\rightarrow& QA Q^\dagger\ \ ,\nonumber\\
B&\rightarrow& QB Q^\dagger\ \ ,\nonumber\\
s&\rightarrow& Qs R^\dagger\ \ ,\nonumber\\
t&\rightarrow& Rt Q^\dagger\ \ ,
\eeqa
with $Q\in U(k), R\in U(N)$ leave $\omega_{\bC}, \omega_{\bR}$ invariant,
and are the analogous of (\ref{nonbanane}).
Let $\xi$ be a generator of the algebra which leaves $\omega^i$
invariant, 
\beq
\label{liedev}
L_\xi\omega^i=0
\ \ ,
\eeq
where $L_\xi$ is the Lie derivative along $\xi$.
As $\omega^i$ is K\"ahler, (\ref{liedev}) gives rise to conserved
quantities, called momentum maps,  defined as
\beq
i(\xi)\omega^i=d\mu^i_\xi
\ \ ,
\eeq
where $\mu^i_\xi=\mu^i_a\xi^a$; 
in complex notation 
\beqa
\mu_{\bC}&=&[A,B]+st\ \ ,
\nonumber\\
\mu_{\bR}&=&[A,A^\dagger]+[B,B^\dagger]+ss^\dagger-t^\dagger t
\ \ .
\eeqa
$\mu^i_\xi=0$ defines a hypersurface
\beq
\mathscr{N}^{+}=\bigg\{\{A,B,s,t\}=x\in M:\ \ \mu_\xi^i=0\bigg\}
\eeq
of dimension ${\rm dim}\, \mathscr{N}^{+}=k^2+4kN$; using (\ref{salute1}),
one can immediately see that these equations are the equivalent
of (\ref{a.2}).
The moduli space of self--dual gauge
connections, $\mathscr{M}^{+}$, is obtained by 
modding $\mathscr{N}^{+}$ by the reparametrizations  
defined in (\ref{invmod}). It has dimension 
${\rm dim}\,\mathscr{M}^{+}=4kN$ and it is 
hyperk\"ahler.%
\footnote{The metric on $\mathscr{M}^{+}$ 
could also be obtained from the K\"ahler form 
$\omega_{\mathscr{M}^{+}}$, which in turn 
is expressed in terms of the 
K\"ahler potential ${\cal K}$, as \cite{maciocia} 
$\omega_{\mathscr{M}^{+}}=\p\bar{\p}{\cal K}=  
{1\over 2}\p\bar{\p}
\Tr \left[a^\dagger(1+P_\infty)a \right]$, 
where $P_{\infty} = 1 - b b^\dagger$ is the asymptotic expression 
of the projector $P$ defined in (\ref{f.666}).}

In the following, we will focus on the $k=2$ case with gauge group $SU(2)$.  
For the explicit computations we go back to the parametrization 
of the ADHM 
moduli space introduced in sec.\,\ref{derrick} and exploited
for the $k=2$ case in sec.\,\ref{victor}; 
the matrix $a$ is written in (\ref{f.14}). 
We introduce a 20--dimensional hyperk\"{a}hler manifold 
$M=(w_1, w_2, a_3, a_1, x_0)$.%
\footnote{Notice that, since we are
using a different parametrization of the ADHM space with respect to 
(\ref{salute1}), the dimension of the manifolds $M$ and $\mathscr{N}^{+}$ 
is  not that of the previous discussion.
However, also the reparametrization groups are different, in such a way 
that the final dimension of the moduli space of self--dual
gauge connections is the same, as it must be.}
Actually, since the theory is invariant under the group of
translations in $\Rq$, one can  fix  
$x_0$ and restrict the analysis  to 
the  16--dimensional hyperk\"{a}hler manifold
$M \backslash \{x_0\}$
parametrized by the quaternionic coordinates 
$m^I=(w_1,w_2,a_3,a_1)$, 
endowed with a flat metric 
\beq
ds^2 = 
\eta_{I\bar J}dm^I d\bar m^{\bar J} = 
|dw_1|^2 + |dw_2|^2 + |da_3|^2 + |da_1|^2   
\ \ .
\label{piatta}
\eeq
To keep the notation 
as simple as possible,  we rename $\mathscr{M}^+\backslash \{x_0\}$ and
$\mathscr{N}^+\backslash \{x_0\}$ as 
$\mathscr{M}^+$, $\mathscr{N}^+$ respectively. 

For $k=2$, the ADHM bosonic constraint (\ref{a.2}) reads
\beq
\bar w_2 w_1 - \bar  w_1 w_2 = 2(\bar a_3 a_1 - \bar a_1 a_3)
\ \ ,
\label{bosco}
\eeq
and, as discussed in sec.\,\ref{derrick}, it is invariant under 
the reparametrization group $O(2)$, whose action on the $k=2$
quaternionic coordinates is
\beqa
&&(w_1^\theta , w_2^\theta) = (w_1 , w_2) R_\theta \ \ , \nonumber \\
&&(a_3^\theta , a_1^\theta) = (a_3 , a_1) R_{2\theta} \ \ ,
\label{ohdue}
\eeqa
with
\beq
R_\theta =
\pmatrix{\cos\theta & \sin\theta \cr -\sin\theta & \cos\theta} \ \ .
\label{rmatrix}
\eeq 
The construction of the reduced bosonic moduli space $\mathscr{M}^+$ 
proceeds now in two steps.
First, we solve explicitly (\ref{bosco})
in an $O(2)$ invariant way.
Since the constraint (\ref{bos}) corresponds to $3k(k-1)/2$ equations,
$\mathscr{N}^{+}$ 
turns out to be a 13--dimensional manifold for $k=2$, 
described by the set of coordinates $(w_1,w_2,a_3,\Sigma)$, where 
$\Sigma$ is the auxiliary real variable related 
to the $O(2)$ reparametrization symmetry. 
Second, we mod out this isometry group
of $\mathscr{N}^{+}$ by means of the hyperk\"ahler quotient procedure. 
The instanton moduli space is then $\mathscr{M}^{+}=\mathscr{N}^{+}/O(2)$, 
and it has dimension 
${\rm dim}\,\mathscr{M}^{+}={\rm dim}\,\mathscr{N}^{+}- k(k-1)/2|_{k=2}=12$.
As anticipated, the construction of the quotient space 
$\mathscr{M}^{+}$ 
can be seen as the gauging of a non--linear sigma model. 
The corresponding connection is given by \cite{hklr}
\beq
{\cal C} = {1\over{|k|^2}}\eta_{I\bar J}
    \bigl(\bar k^{\bar J} dm^I + d\bar m^{\bar J}k^I \bigr) \ \ ,
\label{pizza}
\eeq
where $k^I\partial_I + \bar k^{\bar I}\bar\partial_{\bar I}$ 
is the $O(k)$ Killing vector with 
$|k|^2 = \eta_{I\bar J} k^I \bar k^{\bar J}$.
The components of the $O(2)$ Killing vector on $M$
are easily deduced from (\ref{ohdue}): 
\beq
k^I = (-w_2, w_1, -2a_1, 2a_3) \ \ .
\label{killing}
\eeq
Substituting (\ref{killing}) into (\ref{pizza}), we get 
\beqa
{\cal C} &=& 
{1\over H}\Bigl(\bar w_1 dw_2 - \bar w_2 dw_1 + 
2\bar a_3da_1 - 2\bar  a_1da_3 + 
\nonumber\\ 
&& + d\bar w_2 w_1 -d\bar w_1 w_2 + 2d\bar a_1 a_3 -2d\bar a_3 a_1 \Bigr) 
\ \ .
\label{a}
\eeqa
Notice that this is exactly the connection 
(\ref{connk=2}) obtained in sec.\,\ref{victor} 
by solving the fermionic constraint 
(\ref{fconstr}).
Therefore, this procedure clarifies the geometrical meaning 
of the connection ${\cal C}$ introduced in sec.\,\ref{derrick}, 
providing a very simple method to compute it
{\it directly} from the isometries of the ADHM moduli space, 
without referring 
to the constraint equation (\ref{fconstr}).

The metric 
$g^{\mathscr{N}^{+}}_{I\bar J}$
on the constrained hypersurface $\mathscr{N}^{+}$
is obtained plugging (\ref{natale}) into (\ref{piatta}),  
and gets simplified if we introduce the variable 
\beq
W = \bar{w}_2 w_1 \ \ .
\label{wgrande}
\eeq
The hypersurface $\mathscr{N}^+$ is now described by the new set of 
coordinates
$(w_1, U, V, a_3, \Sigma)$, where 
\beqa
&& U = {{W + \overline W}\over 2} \ \ , \nonumber\\
&& V = {{W - \overline W}\over 2} 
\ \ ,
\label{uvu}
\eeqa
are respectively the real and the imaginary part of $W$. 
The Jacobian factor associated to this change of variables 
is 
\beq
d^4w_1 dU d^3V = |w_1|^4 d^4w_1 d^4w_2 
\ \ .
\label{iacobo}
\eeq
In the new variables, (\ref{piatta}) reads
\beqa 
ds^2 
&=& \Bigl(1+{{|w_2|^2}\over{|w_1|^2}}\Bigr)|dw_1|^2 + {dU^2\over{|w_1|^2}} 
   + {|dV|^2\over{|w_1|^2}} + \nonumber\\
&& - {dU\over{|w_1|^2}} (\bar w_2 dw_1 + d\bar w_1 w_2) 
   + {dV\over{|w_1|^2}} (\bar w_2 dw_1 - d\bar w_1 w_2) + \nonumber\\
&& + |da_3|^2 + |da_1|^2 \ \ ,
\label{flat}
\eeqa
which, inserting (\ref{natale}), becomes 
\beqa 
ds^2 &=& 
\Bigl(1+{{|w_2|^2}\over{|w_1|^2}}\Bigr)|dw_1|^2 + {dU^2\over{|w_1|^2}} 
   + {|dV|^2\over{|w_1|^2}} + \nonumber\\
&& - {dU\over{|w_1|^2}} (\bar w_2 dw_1 + d\bar w_1 w_2) 
   + {dV\over{|w_1|^2}} (\bar w_2 dw_1 - d\bar w_1 w_2) + \nonumber\\
&& + \Bigl(1+{{|a_1|^2}\over{|a_3|^2}}\Bigr)|da_3|^2      
   + {{d\Sigma}^2\over{16|a_3|^2}}
   + {|dV|^2\over{4|a_3|^2}} + \nonumber\\
&& - {{d\Sigma}\over{4|a_3|^2}} (\bar a_1 da_3 + d\bar a_3 a_1)
   - {dV\over{2|a_3|^2}} (\bar a_1 da_3 - d\bar a_3 a_1) 
\ \ .
\label{come} 
\eeqa
The r.h.s. of (\ref{come}) can be regarded as the Lagrangian density of a 
zero--dimensional non--linear sigma model with target space 
$\mathscr{N}^{+}$.
In real coordinates $m^A=(w_1^\mu, U, V^i, a_3^\mu, \Sigma)$, 
the $O(2)$ Killing vector on this manifold has components
\beq
k^A = \Bigl(-w_2^\mu, |w_1|^2 - |w_2|^2, 0, -2a_1^\mu, 
8(|a_3|^2 - |a_1|^2)\Bigr) \ \ .
\label{kill}
\eeq
The global $O(2)$ symmetry can be promoted to a 
local one by introducing the connection (\ref{pizza}), 
which on $\mathscr{N}^{+}$ is written as 
\beqa
{\cal C} &=&  {{g_{AB}^{\mathscr{N}^+}k^B}\over H}dm^A = \nonumber\\
&=& {1\over H}\Bigl(- 2w_2^\mu dw_1^\mu + dU - 4a_1^\mu da_3^\mu +
{{d\Sigma}\over 2}\Bigr) \ \ ,
\label{pizza1}
\eeqa
where the metric  
$g_{AB}^{\mathscr{N}^{+}}$ is obtained by rewriting (\ref{come})
in the coordinates $\{m_A\}$.
Writing $U$ in terms of $w_1, w_2$ by means of (\ref{wgrande}) 
and (\ref{uvu}),
the connection (\ref{pizza1}) becomes 
\beq
{\cal C} = 
{1\over H}\Bigl( w_1^\mu dw_2^\mu -w_2^\mu dw_1^\mu - 4a_1^\mu da_3^\mu + 
{{d\Sigma}\over 2}\Bigr) \ \ .
\label{aripizza}
\eeq 
From the 
gauged version of the Lagrangian (\ref{come}) 
we can read off the metric on $\mathscr{M}^{+} = 
\mathscr{N}^{+}/O(2)$ written in the $\{m^A\}$ coordinates, 
namely \cite{hklr}
\beq
g^{\mathscr{M}^{+}}_{AB}= g^{\mathscr{N}^{+}}_{AB} - 
{{g_{AC}^{\mathscr{N}^{+}}g_{BD}^{\mathscr{N}^{+}}k^Ck^D}
\over {g_{EF}^{\mathscr{N}^{+}}k^Ek^F}} 
\ \ .
\label{metric}
\eeq 
The local $O(2)$ isometry allows one to put $\Sigma$ to zero;
notice that in this gauge (\ref{aripizza}) leads to  
the connection (\ref{connk=2/2}).
Finally, by using translational invariance to restore the 
dependence on $x_0$, and
taking into account the Jacobian factor (\ref{iacobo}), we 
write the volume form on the moduli space of self--dual gauge connections 
with winding number $k=2$ as  
\beq
|w_1|^4\sqrt{g^{\mathscr{M}^{+}}_{\Sigma=0}}
d^4w_1d^4w_2d^4a_3d^4x_0 =
{H\over{|a_3|^4}}\Big| |a_3|^2 - |a_1|^2 \Big| d^4w_1d^4w_2d^4a_3d^4x_0 
\ \ ,
\label{fine}
\eeq
which reproduces the well--known result of Osborn \cite{osb}.



\section*{Acknowledgements}
We are particularly indebted to G.C. Rossi for many valuable
discussions over a long time, 
and for enlightening comments and suggestions 
on a preliminary version of this paper.
We are also grateful to 
D. Anselmi, C.M. Becchi, P. Di Vecchia, S. Giusto,
C. Imbimbo,  V.V. Khoze, M. Matone, 
S.P. Sorella and R. Stora for many stimulating conversations.
D.B. was partly supported by the Angelo Della Riccia Foundation.

\end{document}